%% file: gRouting.tex
\documentclass[sigconf,edbt]{acmart-edbt2018}

\input{macros}

\usepackage{booktabs} 

\setcopyright{rightsretained}

\acmDOI{}

\acmISBN{978-3-89318-078-3}

\begin{document}
\title{On Smart Query Routing: For Distributed \\ Graph Querying with Decoupled Storage}
\author{%
{Arijit Khan $\quad$  Gustavo Segovia $\quad$ Donald Kossmann}%
\vspace{1.6mm}\\
\fontsize{10}{10}\selectfont\itshape
NTU Singapore $\quad$ ETH Zurich, Switzerland $\quad$ Microsoft Research, Redmond, USA\\
\fontsize{9}{9}\selectfont\ttfamily\upshape
arijit.khan@ntu.edu.sg $\quad$ gsegovia@student.ethz.ch $\quad$ donaldk@microsoft.com
}

\input{abstract}

\maketitle

\input{intro}

\input{system}
\input{routing}

\input{experiments}

\input{related}

\input{conclusion}

\vspace{-1mm}
{\scriptsize
\bibliographystyle{ACM-Reference-Format}
\bibliography{ref}
}

\end{document}

%% file: macros.tex
\usepackage{varwidth}
\usepackage{amsmath}
\usepackage{amssymb}
\usepackage{amsfonts}
\usepackage{epsfig}
\usepackage{graphicx}
\usepackage{subfigure}
\usepackage{url}
\usepackage[bottom]{footmisc}
\usepackage{balance}
\usepackage{algorithmic}
\usepackage{algorithm}
\usepackage{multirow}
\usepackage{xspace}
\usepackage[singlelinecheck=off]{caption}
\DeclareCaptionType{copyrightbox}

\newtheorem{requirement}{Requirement}

\newcommand{\spara}[1]{\smallskip\noindent{\bf #1}}

\newcommand{\squishlist}{
 \begin{list}{$\bullet$}
  {  \setlength{\itemsep}{0pt}
     \setlength{\parsep}{3pt}
     \setlength{\topsep}{3pt}
     \setlength{\partopsep}{0pt}
     \setlength{\leftmargin}{2em}
     \setlength{\labelwidth}{1.5em}
     \setlength{\labelsep}{0.5em}
} }
\newcommand{\squishlisttight}{
 \begin{list}{$\bullet$}
  { \setlength{\itemsep}{0pt}
    \setlength{\parsep}{0pt}
    \setlength{\topsep}{0pt}
    \setlength{\partopsep}{0pt}
    \setlength{\leftmargin}{2em}
    \setlength{\labelwidth}{1.5em}
    \setlength{\labelsep}{0.5em}
} }

\newcommand{\squishdesc}{
 \begin{list}{}
  {  \setlength{\itemsep}{0pt}
     \setlength{\parsep}{3pt}
     \setlength{\topsep}{3pt}
     \setlength{\partopsep}{0pt}
     \setlength{\leftmargin}{1em}
     \setlength{\labelwidth}{1.5em}
     \setlength{\labelsep}{0.5em}
} }

\newcommand{\squishend}{
  \end{list}
}

\newcommand{\eat}[1]{}

\newcommand{\NP}{\ensuremath{\mathbf{NP}}\xspace}
\newcommand{\onlyP}{\ensuremath{\mathbf{P}}\xspace}

\newcounter{ccc}

\newcommand{\bigO}{\mathcal{O}}

%% file: abstract.tex
\begin{abstract}
We study online graph queries that retrieve nearby nodes of a query node in a large network.
To answer such queries with high throughput and low latency, we partition the graph and process in parallel across a cluster of servers.
Existing distributed graph systems place each partition on a separate server, where query answering over that partition takes place.
This design has two major disadvantages. First, the router maintains a fixed routing table (or, policy). Hence, these systems are less flexible
for query routing, fault tolerance, and graph updates. Second, the graph must be partitioned such that the workload across servers is
balanced, and the inter-machine communication is minimized. In addition, to maintain good-quality partitions, it is required to update the existing partitions
based on workload changes. However, graph partitioning, online monitoring of workloads, and dynamically updating the partitions are expensive.

In this work, we mitigate both these problems by decoupling graph storage from query processors, and by developing smart routing strategies with the usage of
graph landmarks and embedding, that improve cache locality in query processors. Since a query processor is no longer assigned any fixed part of the graph,
it is equally capable of handling any request, thus facilitating load balancing and fault tolerance. On the other hand, due to our smart routing strategies,
query processors can effectively leverage their cache, reducing the impact of how the graph is partitioned across storage servers. A detailed
empirical evaluation with several real-world, large graph datasets demonstrates that our proposed framework {\sf gRouting}, even with simple hash
partitioning, achieves up to an order of magnitude better query throughput compared to existing distributed graph systems that employ expensive
graph partitioning and re-partitioning strategies.
\end{abstract}

%% file: intro.tex
\vspace{-1mm}
\section{INTRODUCTION}
\label{sec:intro}
Graphs with millions of nodes and billions of edges are ubiquitous to represent highly interconnected
structures including the World Wide Web, social networks,
knowledge graphs, genome and scientific databases, Internet of things, medical and government records.
In order to support online search and query services (possibly from many clients) with low latency and high throughput,
data centers and cloud operators consider scale-out solutions, in which the graph and its data are partitioned horizontally
across cheap commodity servers in the cluster. We assume that the graph topology and the data associated with nodes and edges are co-located,
since they are often accessed together \cite{WXSXHZ13, MD12, MD14}. Keeping with the modern database trends to support low-latency
operations, we target a fully in-memory system, and use disks only for durability \cite{YYZK12, SEHM13, SWL13}.
\begin{figure}[t!]
\centering
\includegraphics[scale=0.29]{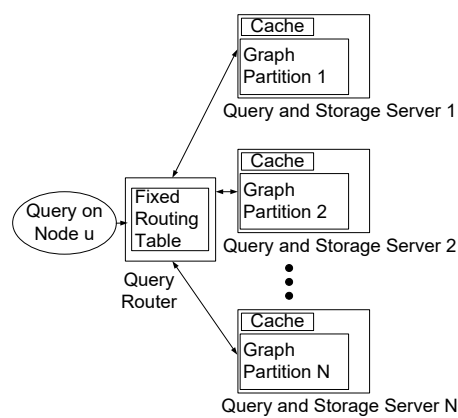}
\vspace{-2mm}
\caption{\small State-of-the-art distributed graph querying systems (e.g., {\sf SEDGE} \cite{YYZK12}, {\sf Trinity} \cite{SWL13}, {\sf Horton} \cite{SEHM13})}
\label{fig:existingsystem}
\vspace{-6mm}
\end{figure}

In this paper, we study online queries that explore a small region of the entire graph, and
require fast response time. These queries usually start with a query node, and traverse its neighboring nodes
up to a certain number of hops. While we shall formally introduce our queries in Section~\ref{sec:system},
typical real-world examples can be as follows. {\bf (1)} {\em Home Timeline Query} \cite{WXSXHZ13}:
When a user logs in to {\sf Facebook} or refreshes her homepage, a timeline is shown with recent
posts created by her friends. {\bf (2)} {\em Ego-Centric Query} \cite{MD14}: In the {\sf LinkedIn} social
network, user Alice may search for her connections within 2-hops who are currently employed by {\sf Google}.
{\bf (3)} {\em Pattern Matching Query} \cite{MD12}: Given the {\sf Microsoft} Academic graph, find all papers on
``distributed graph systems'' which are a result of collaboration between researchers from {\sf UC Berkeley} and {\sf CMU}.
We generalize such online queries with an $h$-hop traversal
from the query node(s).

For efficient answering of online $h$-hop queries in a distributed environment, state-of-the-art systems (e.g., \cite{YYZK12,SWL13, SEHM13})
first partition the graph, and then place each partition on a separate server, where query answering over that partition takes place
(Figure~\ref{fig:existingsystem}). Since the server which contains the query node can only handle that request, the router
maintains a fixed routing table (or, a fixed routing strategy, e.g., modulo hashing).
Hence, these systems are less flexible with respect to query routing and fault tolerance, e.g., adding more
machines will require updating the routing table. Besides, an effective graph
partitioning in these systems must achieve: (1) workload balancing to maximize parallelism, and (2) locality of data access to
minimize network communication. It has been shown in the literature \cite{YYZK12} that
sophisticated partitioning schemes improve the performance of graph querying, compared to an inexpensive hash partitioning.

However, due to power-law degree distribution of real-world graphs, it is difficult to get high-quality
partitions \cite{GLGBG12}.
Besides, a one-time partitioning cannot cope with later updates to graph
structure or variations in query workloads. Several graph re- partitioning and replication-based strategies were proposed to alleviate these problems,
e.g., \cite{YYZK12,NKDC15,MD12}. However, online monitoring of workload changes, re-partitioning of the graph topology, and migration of graph data across servers are expensive;
and they reduce the efficiency and throughput of online querying \cite{RBMZ15}.
%

\begin{figure}[t!]
\centering
\includegraphics[scale=0.3]{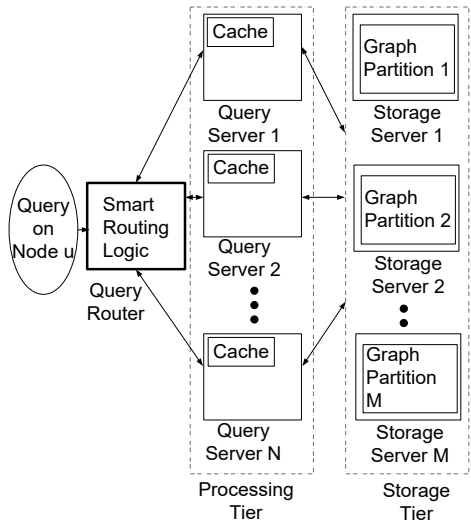}
\vspace{-2mm}
\caption{\small Decoupled architecture for distributed graph querying}
\label{fig:oursystem}
\vspace{-6mm}
\end{figure}

\spara{Our Contribution.} In contrast to existing systems, we consider a
different architecture, which relies {\em less}
on an effective graph partitioning. Instead, we decouple query processing and graph storage
into two separate tiers (Figure~\ref{fig:oursystem}).
In a decoupled framework, the graph is partitioned across servers allocated to the storage tier,
and these storage servers hold the graph data in their main memory. Since a query processor is no longer assigned any fixed part of the graph, it is equally
capable of handling any request, thus facilitating load balancing and fault tolerance. At the same time,
the query router can send a request to any of the query processors, which adds more flexibility to query routing, e.g., more query processors can be added (or, a query processor
that is down can be replaced) without affecting the routing strategy.
Another benefit due to decoupled design is that each tier can be
scaled-up independently. If a certain workload is processing intensive, more servers could be allocated to the processing tier. On the contrary, if the graph size increases over time, more
servers can be added in the storage tier.  This decoupled architecture, being generic, can be employed in many existing graph querying systems.

The idea of decoupling, though effective, is not novel.
{\sf Facebook} implemented a fast caching layer, {\sf Memcached} on top of a graph database that
scales the performance of graph query answering \cite{NFGKLLMPPSSV13}. {\sf Google}'s {\sf F1} \cite{SVSH13}
and {\sf ScaleDB} (http://scaledb.com/ pdfs/TechnicalOverview.pdf) are based on a decoupling principle for scalability.
Recently,  Loesing et. al. \cite{LPEK15} and  Binnig et. al. \cite{BCGKZ16} demonstrated the benefits of a decoupled, shared-data architecture, together with low
latency and high throughput Infiniband network. Shalita et. al. \cite{SKKSPAKS16} employed de-coupling for an optimal assignment of
HTTP requests over a distributed graph storage.

{\em Our fundamental contribution lies in designing a smart query routing logic to utilize the cache of query processors over such decoupled architecture}.
Achieving more cache hits is critical in a decoupled architecture -- otherwise, the query processors need to retrieve the data from storage
servers, which will incur extra communication costs. However, this is a non-trivial problem, e.g., exploiting cache locality and balancing
workloads are conflicting in nature. As an example, to achieve maximum cache locality, the router can send
all the queries to the same processor (assuming no cache eviction
happens). However, the workload of the processors will be
highly imbalanced in that case, resulting in lower throughput.
In addition, graph querying workloads are
significantly different from traditional database applications. The interconnected nature of graph data results in
poor locality, and each query usually accesses multiple neighboring nodes spreading across the distributed storage. Therefore, to
maximize cache hit rates at query processors, it is not sufficient to only route the queries on same nodes to the same processor.
Rather, successive queries on {\em neighboring} nodes should also be routed to the same processor, since the neighborhoods
of two nearby nodes may significantly overlap. Besides, the smart routing logic that we design to resolve the above challenges,
should be lightweight on storage and online computation cost, as well as adaptive with workload hostspot and graph updates.
To the best of our knowledge, such smart query routing schemes for effectively leveraging the cache contents were not considered
in state-of-the-art graph querying systems, including {\sf SEDGE} \cite{YYZK12}, {\sf Trinity} \cite{SWL13}, {\sf Horton} \cite{SEHM13}.
{\em In this paper, we take the first step to this goal}.

\vspace{0.8mm}

We summarize our contributions as follows.
\begin{enumerate}
\item We investigate for the first time the novel problem of {\em smart query routing} aimed at improving the throughput and efficiency of distributed graph querying.
\item In contrast to many distributed graph querying systems \cite{YYZK12,SWL13, SEHM13}, we consider a different architecture that {\em decouples} query
processors from storage layer, thereby achieving flexibility in system deployment, query routing, scaling up, load balancing, and fault tolerance.
\item We develop smart, lightweight, and adaptive {\em query routing algorithms} that improve cache hit rates at query processors, thus reducing communication
with storage layer, and making our design less reliant on a sophisticated graph partitioning scheme across storage layer.
\item We empirically demonstrate throughput and efficiency of our framework, {\sf gRouting} on four real-life graphs, while also comparing with two state-of-the-art distributed
graph processing systems ({\sf SEDGE/Giraph} \cite{YYZK12} and {\sf PowerGraph} \cite{GLGBG12}).
Our decoupled implementation, even with its simple hash partitioning,
achieves up to an order of magnitude higher throughput compared to these existing systems with expensive
graph partitioning and re-partitioning schemes.
\end{enumerate}

%% file: system.tex
\vspace{-1mm}
\section{PRELIMINARIES}
\label{sec:system}
%
\subsection{Graph Data Model}
\label{sec:graph_data}
A heterogeneous network can be modeled as a
labeled, directed graph $G = (V,E,\mathcal{L})$ with node set $V$, edge set $E$, and
label set $L$, where (1) each node $u \in V$ represents an entity in the network,
(2) each directed edge $e \in E$ denotes the relationship between two entities,
and (3) $\mathcal{L}$ is a function which assigns to each node $u$ and every edge
$e$ a label $\mathcal{L}(u)$ (and $\mathcal{L}(e)$, respectively) from a finite alphabet.
In practice the node labels may represent the attributes of
the entities, such as name, job, location, etc, and edge labels the
type of relationships, e.g., founder, place founded, etc.

\begin{figure}[tb!]
\centering
\subfigure [\small Graph data]{
\includegraphics[scale=0.19]{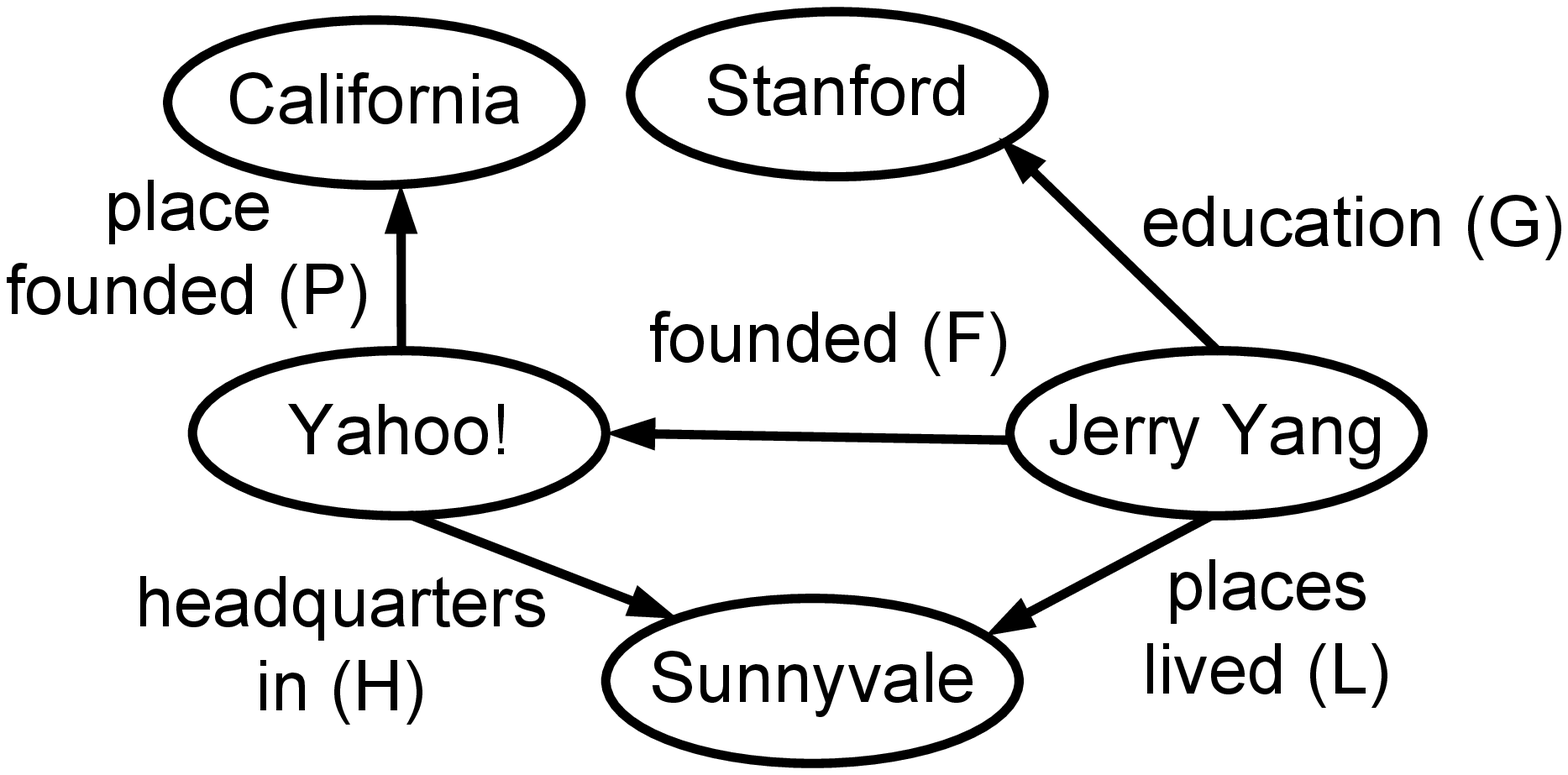}
\label{fig:key_value}
}
\subfigure [\small Graph in key-value format] {
\includegraphics[scale=0.19]{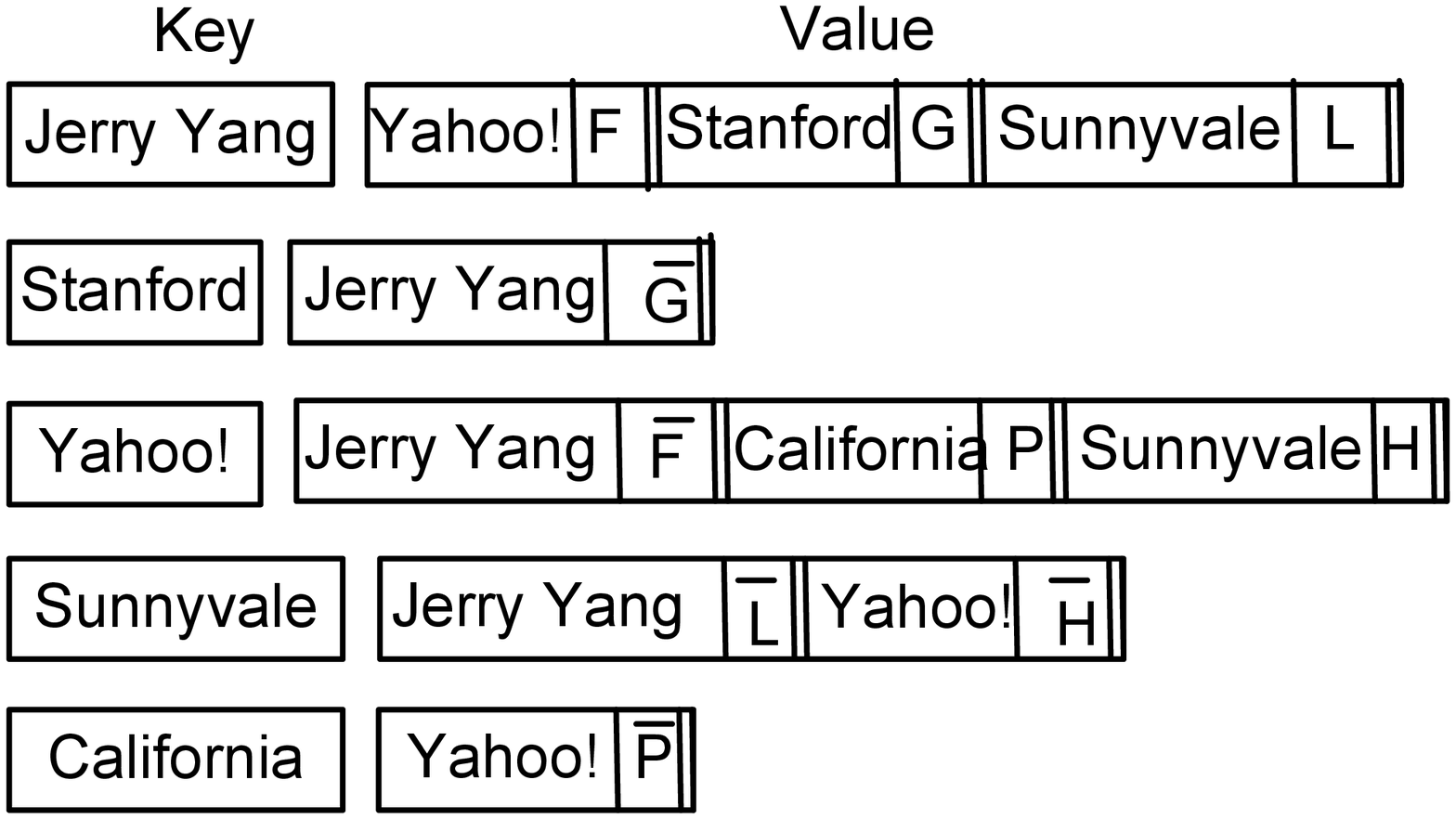}
\label{fig:kv2}
}
\vspace{-4mm}
\caption{\small  Key-Value storage of graph: $\bar{F}$ is inverse relationship of $F$.}
\label{fig:keyvalue}
\vspace{-5mm}
\end{figure}

We store the graph
as an adjacency list (Figure~\ref{fig:keyvalue}). Every node in the graph is added as an entry in
the storage where the key is the node id and the value is an array of 1-hop
neighbors. If the nodes and edges have labels, they are stored
in the corresponding value entry.
For each node, we store both incoming and outgoing edges. This is because both incoming and outgoing edges of a node can be important
from the context of different queries. As an example, if there is an edge {\em founded} from {\em Jerry Yang} to {\em Yahoo!}
in some knowledge graph, there also exists a reverse relation {\em founded\_by} from {\em Yahoo!} to {\em Jerry Yang}. Such
information could be useful in answering queries about {\em Yahoo!}.
\vspace{-1mm}
\subsection{h-Hop Traversal Queries}
\label{sec:query}
We introduce various $h$-hop queries over heterogeneous, directed graphs in the following.
\begin{enumerate}
\item {\bf $h$-hop Neighbor Aggregation}: Count the number of $h$-hop neighbors (or, number of occurrences of a specific label within $h$-hop neighborhood) of a query node.
\item {\bf $h$-step Random Walk with Restart}: The query starts at a node, and runs for $h$-steps --- at each successive step jumps to one of its neighbors with equal probability, or returns to the query node with a small probability.
\item {\bf $h$-hop Reachability}: Find if a given target node is reachable from a given source node within $h$-hops.
\end{enumerate}

The aforementioned queries are often used as the basis for more complex graph operations. For example,
neighborhood aggregation is critical for node labeling and classification, that is,
the label of an unlabeled node could be assigned as the most frequent label which is present within its $h$-hop neighborhood.
The $h$-step random walk is useful in expert finding, ranking, discovering functional modules, complexes, and pathways.
For processing the third query, we perform a bi-directional
breadth first search (BFS), i.e., in forward direction from the source node and in backward direction from the target node,
which is feasible because we store both incoming and outgoing edges for every node in the graph dataset. In addition,
if there are node- and edge-label constraints in reachability computation, one can enforce such constraints while performing
the BFS. Therefore, our third query can be employed
in distance-constrained and label-constrained reachability search, as well as in
approximate graph pattern matching queries \cite{MD12}.

\vspace{-2mm}
\subsection{Decoupled Design}

We review the general implementation details
of our decoupled architecture ---
discussing their merits, and how to overcome their limitations,
which will lead to introducing our smart query routing algorithms in the
following section.


We decouple query processing from graph storage. This decoupling happens at a {\em logical} level.
As an example, query processors can be different physical machines
than storage servers. On the other hand, the same physical machine can also run a query processing daemon, together with
storing a graph partition in its main memory as a storage server. However, the logical separation between the two layers
is critical in our design.

The advantages of this separation are more flexibility in query routing, system
deployment, and scaling up, as well as achieving better load balancing and fault tolerance.  However, we must also
consider the drawbacks of having the graph storage apart. First,
query processors may need to communicate with the storage tier via the network.
This includes an additional penalty to the response time for answering a query.
Second, it is possible that this design causes high contention rates on either the
network, storage tier, or both.

To mitigate these issues, we design smart routing strategies that route online queries to processors which are likely to have the relevant data in their cache, thereby reducing the communication overhead between processing and storage tiers. Below, we discuss various components of our design, including storage, processing tier, and router.

\spara{Graph Storage Tier.} The storage tier holds all graph data by horizontally partitioning it
across cheap commodity servers. A sophisticated graph partitioning scheme could benefit our decoupled architecture
in the following way. Let us assume that the neighboring nodes can be stored in a page
within the same storage server, and the granularity of transfer from storage to processing tier is a page containing several nodes.
Then, we could actually ship a set of relevant nodes with a single request if the graph is partitioned well. This will
reduce the number of times data are transferred between the processing and storage tier.

However, our lightweight and smart query routing techniques exploit the notion of graph landmarks \cite{KSW09} and embedding \cite{ZSWZZ10},
thereby effectively utilizing the cache of query processors that stores recently used graph data.
As demonstrated in our experiments, due to our smart routing, many neighbors up to 2$\sim$3-hops of a query node can be found locally in the query processors' cache.
Therefore, the partitioning scheme employed across storage servers becomes less important.

\spara{Query Processing Tier.} The query processing tier consists of servers where the actual query processing takes place.
These servers do not communicate with each other \cite{LPEK15}. They only receive queries
from the query router. They can also request graph data from the storage tier, if necessary.

To reduce the amount of calls made to the storage tier, we utilize the cache of the query processors.
Whenever some data is retrieved from the storage, it is saved in cache, so that the same request can be avoided in the near future.
However, it imposes
a constraint on the maximum storage capacity. When the addition of a new entry surpasses this storage limit, one
or more old entries are evicted from the cache. We chose the {\sf LRU} (i.e., Least Recently Used) eviction
policy because of its simplicity. {\sf LRU} is usually implemented as the default cache replacement policy, and it
favors recent queries. Thus, it performs well with our smart routing schemes.

\spara{Query Router.} The router creates a thread for each processor and opens a connection to send queries
by following the routing schemes that we shall describe in Section~\ref{sec:routing}.

%% file: routing.tex
\vspace{-1mm}
\section{QUERY ROUTING STRATEGIES}
\label{sec:routing}
When a query arrives at the router, the router decides the appropriate query processor to which the request could be sent.
For state-of-the-art graph querying systems, e.g., {\sf SEDGE} \cite{YYZK12} and {\sf Horton} \cite{SEHM13}, where each query processor is assigned a graph partition,
this decision is fixed and defined in the routing table; the processor which contains the query node should handle the request. With a decoupled
architecture, no such
mapping exists. Hence, we design novel routing schemes with the following objectives.
\vspace{-1mm}
\subsection{Routing Algorithm Objectives}
We aim at improving the query throughput and reducing the latency. To this end, we identify the following
criteria that a {\em smart routing} scheme must possess.

\spara{1. Leverage each processor's cached data.} To formalize this notion, let us consider $t$ queries $q_1, q_2, \ldots, q_t$ received successively
by the query router. The router will send them to query processors in a way such that the average number of {\em cache hits} at the processors is maximized.
This, in turn, reduces the average query processing latency. However, as stated earlier, to achieve maximum cache hits,
it will not be sufficient to only route the queries on same nodes to the same processor. Rather, successive queries on {\em neighboring}
nodes should also be routed to the same processor, since the neighborhoods of two nearby nodes may significantly overlap. This will be discussed shortly in
Requirement 1.

\spara{2. Balance workload even if skewed or contains hotspot.} As earlier, let us consider a set of $t$ successive queries. A  na\"{\i}ve approach
will be to ensure that each query processor receives equal number of queries, e.g., a round-robin way of query dispatching by the router. However, each
query might have a different workload, and would require a different processing time. We, therefore, aim at maximizing the overall {\em throughput}
via query stealing (explained in Requirement 2), which automatically balances the workload across query processors.

\spara{3. Make fast routing decisions.} The average time at the router to dispatch a query should be minimized, ideally a small constant time, or
much smaller than $\bigO(n)$, where $n$ is the number of nodes in the input graph. This reduces the query processing latency.
\begin{figure}[t!]
\centering
\includegraphics[scale=0.39]{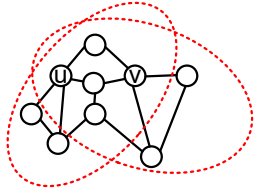}
\vspace{-2mm}
\caption{\small Topology-aware locality: 2-hop neighborhoods of nearby nodes $u$ and $v$ significantly overlap.}
\label{fig:topology_aware}
\vspace{-5mm}
\end{figure}

\spara{4. Have low storage overhead in the router.} The router may store some auxiliary data structures to enable fast routing decisions. However, this additional storage overhead must be a small fraction compared to the input graph size.
\vspace{-1mm}
\subsection{Challenges in Query Routing}
It is important to note that our routing objectives are not in harmony; in fact, they are often conflicting with each other.
First, in order to achieve maximum cache locality, the router can send all the queries to the same processor (assuming no cache eviction happens).
However, the workload of the processors will be highly imbalanced in this case, resulting in lower throughput.
Second, the router could inspect the cache of each processor before making a good routing decision, but this will add
network communication delay. Hence, the router must {\em infer} what is likely to be in each processor's cache.

In the following, we introduce two concepts that are directly related to our routing objectives, and will be useful in
designing smart routing algorithms.

\spara{Topology-Aware Locality.} To understand the notion of cache locality for graph queries (i.e., routing objective 1),
we define a concept called {\em topology-aware locality}.
If $u$ and $v$ are nearby nodes, then successive queries on $u$ and $v$ must be sent to the same processor.
It is very likely that the $h$-hop neighborhoods of $u$ and $v$ will significantly overlap (Figure~\ref{fig:topology_aware}).
We refer to this phenomenon as the topology-aware locality.

But, how will the router know that $u$ and $v$ are nearby nodes? One option is to store the entire graph topology in the router; but this could have a
high storage overhead. For example, the {\em WebGraph} dataset that we experimented with has a topology of size 60GB.
Ideally, a graph with $10^7$ nodes can have up to $10^{14}$ edges, and in such cases, storing only the topology itself requires petabytes of memory.
Thus, we impose a requirement on our smart routing schemes as follows.
\begin{requirement}
The additional storage at the router for enabling smart routing should not be asymptotically larger than $\bigO(n)$, $n$ being the number of nodes;
however, the routing schemes should still be able to exploit topology-aware locality.
\end{requirement}

One may realize that achieving this goal is non-trivial, as the topology size can be $\bigO(n^2)$, and we provision for only $\bigO(n)$ space to
approximately preserve such information.

\spara{Query Stealing.} Routing queries to processors that have the most useful cache data might not always be the best strategy.
Due to power-law degree distribution of real-world graphs, processing queries on different nodes might require different amount of time. Therefore, the processors dealing
with high-degree nodes will have more workloads. Load imbalance can also happen if queries are concentrated in one specific region of the graph.
When that happens, all queries will be sent to one processor, while other processors remain idle. To rectify such scenarios, we implement {\em query stealing} in our routing schemes
as stated next.
\begin{requirement}
Whenever a processor is idle and is ready to handle a new query, if it does not have any other requests assigned to it, it may ``steal'' a request that was originally intended for
another processor.
\end{requirement}

Query stealing is a well established technique for load balancing that is prevalently used by the HPC community, and
there are several ways how one can implement it. The idle processors can actively steal queries from other
processors that are currently busy. In our implementation, we assume that the processors send acknowledgements to the
router when they finish computation. Thus, {\em we perform query stealing at the router level}.
In particular, the router sends the next query to a processor only when it receives an acknowledgement for the previous query
from that processor. The router also keeps a {\em queue} for each connection in order to store the future queries that need to be
delivered to the corresponding processor. By monitoring the length of these queues, it can estimate how busy a processor is, and this
enables the router to rearrange the future queries for load balancing. We demonstrate the effectiveness of query stealing in our
experiments (Section~\ref{sec:sensitivity}).

We next design four different routing schemes --- the first two are na\"{\i}ve and do not meet all the
objectives of smart routing. On the other hand, the last two algorithms follow the
requirements of a smart routing strategy. We, therefore, use the first two routing schemes as baselines
to compare against our smart routing algorithms in the experiments.
\vspace{-2mm}
\subsection{Baseline Methods}
\subsubsection{Next Ready Routing}
{\em Next Ready} routing is our first baseline strategy.
The router decides where to send a query by choosing the next processor that has finished computing and is ready for a new request.
The main advantages are:
\begin{itemize}
\item It is easy to implement.
\item Routing decisions are made in constant time.
\item No preprocessing or storage overhead is required.
\item The workload is well balanced.
\end{itemize}
Even though this scheme has several benefits, it fails to leverage processors' cache.
\subsubsection{Hash Routing}
The second routing scheme that we implement is {\em hash}, and it also serves
as a baseline to compare against our smart routing techniques. The router applies a
fixed hash function on each query node's id to determine the processor
where it sends the request. In our implementation, we apply a modulo
hash function:
\begin{align}
&\text{Target-Processor-Id} \nonumber & \\
&= (\text{Query-Node-Id}) {\sf MOD} (\text{Number-Of-Processors}) &
\end{align}

In order to facilitate load balancing in the presence of workload skew, we implement {\em query
stealing} mechanism. Whenever a processor is idle and is ready to handle a new query, if it does not have any
other requests assigned to it, it steals a request that was originally intended for another processor.
Since queries are queued in the router, the router is able to take this decision,
and ensures that there are no idle processors when there is still some work to be done. Our hash routing has all the benefits
of next ready, and very likely it sends a repeated query to the same processor, thereby
getting better locality out of the cache. However, hash routing cannot capture topology-aware locality.
\vspace{-1mm}
\subsection{Proposed Methods}
\label{sec:proposed}
\subsubsection{Landmark Routing}
\begin{figure}[t!]
\centering
\includegraphics[scale=0.22]{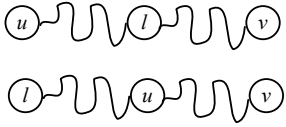}
\vspace{-2mm}
\caption{\small Strict upper (above) and lower (below) bounds on shortest-path distances based on landmark}
\label{fig:landmark}
\vspace{-5mm}
\end{figure}

Our first smart routing scheme is based on the notion of landmark nodes \cite{KSW09}.
One may recall that we store both incoming and outgoing edges of every node, thus we consider
a {\em bi-directed} version of the input graph in our smart routing algorithms. In particular, we assume a
bi-directed edge corresponding to every directed edge in the input graph, then we select
a small set $L$ of nodes as landmarks, and also pre-compute the distance of every node to these landmarks.
We determine the optimal number of landmarks based
on empirical results. As shown in Figure~\ref{fig:landmark}, given some landmark node $l \in L$,
the distance $d(u,v)$ between any two nodes $u$ and $v$ are bounded as follows:
\vspace{-1mm}
\begin{align}
&d(u, v) \le d(u, l) + d(l, v) \nonumber & \\
&d(u, v) \ge |d(u, l) - d(l, v)| &
\end{align}

\vspace{-2mm}
Intuitively, if two nodes are close to a given landmark, they are likely to be close themselves. Our landmark routing is based on the above principle.
We first select a set of landmarks that partitions the graph into $P$ regions, where $P$ is the total number of processors. We then decide a one-to-one
mapping between those regions and processors. Now, if a query belongs to a specific region (decided based on its distance to landmarks), it is routed
to the corresponding processor. Clearly, this routing strategy requires a preprocessing phase,
where we select the landmark nodes.

\spara{Preprocessing.}
We select landmarks based on their node degree and how well they spread over the entire graph \cite{AIY13}. Our first step is to find a certain
number of landmarks considering the highest degree nodes, and then compute their distance to every node in the graph by performing breadth first searches (BFS).
In this process, if we find two landmarks
to be closer than a pre-defined threshold, the one with the lower degree is discarded. The complexity of this step is $\bigO(|L|e)$, due to $|L|$ number of BFS,
where $|L|$ is the number of landmarks, and $e$ is the number of edges.

Next, we assign the landmarks to query processors as follows. First, every processor is assigned a ``pivot'' landmark with the intent that pivot
landmarks are as far from each other as possible.
\begin{itemize}
\item The first two pivot landmarks are the two that are farthest apart considering all other landmark pairs.
\item Each next pivot is selected as the landmark that is farthest from all previously selected pivot landmarks.
\end{itemize}

Each remaining landmark is assigned to the processor which contains its closest pivot landmark. The complexity of this step is $\bigO(|L|^2+|L|P)$, where $P$ is the number
of processors.

Finally, we define a ``distance'' metric $d$ between the graph nodes and query processors. The distance of a node $u$ to a processor $p$ is defined
as the minimum distance of $u$ to any landmark that is assigned to processor $p$. Thus, we compute a distance value $d(u,p)$ for every node $u$ to every processor $p$.
This information is stored in the router, which requires $\bigO(nP)$ space and $\bigO(nL)$ time to compute, where $n$ is the number of nodes. Therefore, {\em the storage requirement at the router is linear in the number of nodes}.

\spara{Routing.}
To decide where to send a query on node $u$, the router verifies the pre-computed distance $d(u,p)$ for every processor $p$, and selects the one with the smallest $d(u,p)$ value.
As a consequence, the routing decision time is linear in the number of processors: $\bigO(P)$. This is very efficient since the number of processors is small.

In contrast to our earlier baseline routings, this method is able to leverage topology-aware locality. It is likely that query nodes that are in the
neighborhood of each other will have similar distances to the processors; hence, they will be routed in a similar fashion. On the other hand, the landmark routing scheme
is less flexible with respect to addition or removal of processors, since the assignment of landmarks to processors, as well as the distances $d(u,p)$ for every node $u$
and each processor $p$ needs to be recomputed.

The distance metric $d(u,p)$ is useful not only in finding the best processor for a certain query, but it can also be used for load balancing, fault tolerance, dealing with
workload skew, and hotspots. As an example, let us assume that the closest processor for a certain query is very busy, or is currently down.
Since the distance metric gives us distances to all
processors, the router is able to select the second, third, or so on closest processor. This form of load balancing will impact the nearby query nodes in the same way; and therefore,
the modified routing scheme will still be able to capture topology-aware locality. In practice, it can be complex to define exactly when a query should be routed
to its next best query processor. We propose a formula that calculates the {\em load-balanced distance} $d^{LB}(u,p)$ between a query node $u$ and a processor $p$.
\vspace{-2mm}
\begin{align}
d^{LB}(u,p) = d(u,p) + \frac{\text{Processor Load}}{\text{Load Factor}}
\label{equ:lb_land}
\end{align}

Thus, the query is always routed to the processor with the smallest $d^{LB}(u,p)$. The router uses the number of queries in the queue corresponding to a processor
as the measure of its load. The load factor is a tunable parameter, which allows us to decide how much load would result in the query to be routed to another processor. We find
its optimal value empirically.

\spara{Dealing with Graph Updates.}
During addition/ deletion of nodes and edges, one needs to recompute the distances from every node to each of the landmarks. This can be performed efficiently
by keeping an additional {\em shortest-path-tree} data structure \cite{TAGVD11}. However, to avoid the additional space and time complexity of maintaining a shortest-path-tree, we
follow a simpler approach. When a new node $u$ is added, we compute the distance of this node to every landmark, and also its distance $d(u,p)$
to every processor $p$. In case of an edge addition or deletion between two existing nodes, for these two end-nodes and their neighbors up to a certain number of hops (e.g., $2$-hops),
we recompute their distances to every landmark, as well as to every processor. Finally, in case of a node deletion, we handle it by
considering deletion of multiple edges that are incident on it. After a significant number of updates, previously selected landmark
nodes become less effective; thus, we recompute the entire preprocessing step periodically in an off-line manner.
\begin{figure}[t!]
\centering
\includegraphics[scale=0.19]{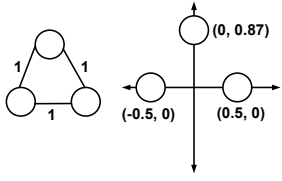}
\vspace{-2mm}
\caption{\small Example of graph embedding in 2D Euclidean plane}
\label{fig:embedding}
\vspace{-5mm}
\end{figure}
%

\subsubsection{Embed Routing}
Our second smart routing scheme is the {\em Embed} routing, which is based on a graph embedding principle \cite{ZSWZZ10,DCKM04}.
We embed a graph into a lower dimensional Euclidean space such that the hop-count distance between graph nodes
are approximately preserved via their Euclidean distance (Figure~\ref{fig:embedding}). As depicted in \cite{ZSWZZ10}
and also found in our experiments, higher dimensions produce smaller errors in terms of preserving
node-pair distances. Particularly, we observed that an embedding of dimensionality 10 or higher is able to preserve nearby distances reasonably well.
We then use the resulting node co-ordinates to determine how far a query node is from the recent history of queries that were sent
to a specific processor. Clearly, embed routing also requires a preprocessing step as discussed in the following.

\spara{Preprocessing.}
For efficiently embedding a large graph in a $D$-dimensional Euclidean plane, we first select a set $L$ of landmarks and find their distances from each node in the graph. We then assign co-ordinates to landmark nodes such that the distance between each pair of landmarks is approximately preserved. We, in fact, minimize the {\em relative error} in distance for each pair of landmarks, defined below.
\vspace{-1mm}
\begin{align}
f_{error}(v_1,v_2) = \frac{|d(v_1,v_2)-\text{EuclideanDist}(v_1,v_2)|}{d(v_1,v_2)}
\end{align}

Here, $d(v_1,v_2)$ is the hop-count distance between $v_1$ and $v_2$ in the original graph, and EuclideanDist$(v_1,v_2)$ is their Euclidean distance after the graph is embedded. We minimize the relative error since we are more interested in preserving the distances between nearby node pairs. Our problem is to minimize the aggregate of such errors over all landmark pairs --- this can be cast as a generic multi-dimensional global minimization problem, and could be approximately solved by many off-the-shelf techniques, e.g., the {\em Simplex Downhill} algorithm that we apply in this work.
Next, every other node's co-ordinates are found also by applying the Simplex Downhill algorithm that minimizes the aggregate relative distance error between the node and all the landmarks.
The overall graph embedding procedure consumes a modest preprocessing time: $\bigO(|L|e)$ due to BFS from $|L|$ landmarks, $\bigO(|L|^2 D)$ for
embedding the landmarks, and $\bigO(n|L| D)$ for embedding the remaining nodes. In addition, the second step is completely parallelizable per node. Since each node receives $D$ co-ordinates, it requires total $\bigO(nD)$ space in the router, which is linear in the number of nodes.

Unlike landmark routing, {\em a benefit of embed routing is that the preprocessing is independent of the system topology, allowing more processors to be easily added at a later time}.

\vspace{-1mm}
\spara{Routing.}
The router has access to each node's co-ordinates. By keeping an
average of the query nodes' co-ordinates that it sent to each processor,
the router is able to infer the cache contents in these processors.
As a consequence, the router can find the distance between a query node $u$ and a
processor $p$, denoted as $d_1(u,p)$, and defined as the
distance of the query node's co-ordinates to the historical mean of the processor's cache contents.
Since recent queries are more likely to influence the cache contents due to {\sf LRU} eviction policy, we use the exponential moving average
function to compute the mean of the processor's cache contents. Initially, the mean co-ordinates for each processor are assigned uniformly
at random. Next, assuming that the last query on node $v$ was sent to processor $p$,
its mean co-ordinates are updated as:
\vspace{-2mm}
\begin{align}
&\text{MeanCo-ordinates}(p) = \alpha\cdot\text{MeanCo-ordinates}(p) \nonumber \\
& \qquad \qquad \qquad \qquad \qquad + (1-\alpha)\cdot \text{Co-ordinates}(v)&
\label{equ:mean}
\end{align}

The smoothing parameter $\alpha \in (0, 1)$ in the above Equation determines the degree of decay used to discard older queries.
For example, $\alpha$ close to $0$ assigns more weight only to the last query, and $\alpha$ close to $1$ decreases the weight
on the last query. We determine the optimal value of $\alpha$ based on experimental results.
Finally, the distance between a query node $u$ and a processor $p$ is computed as given below.
\vspace{-2mm}
\begin{align}
d_1(u,p) = ||\text{MeanCo-ordinates}(p) - \text{Co-ordinates}(u)||
\end{align}

Since we consider embedding in an Euclidean plane, we use the $L_2$ norm to compute distances. We select the
processor with the smallest $d_1(u,p)$ distance. One may observe that the routing decision time is only $\bigO(PD)$, $P$ being the number of
processors and $D$ the number of dimensions.

Analogous to landmark routing, we now have a distance to each processor for a query; and hence, we are able to make routing decisions taking into
account the processors' workloads and faults.
As earlier, we define a {\em load-balanced distance} $d_1^{LB}(u,p)$ between a query node $u$ and a processor $p$,
and the query is always routed to the processor with the smallest $d_1^{LB}(u,p)$ value.
\vspace{-2mm}
\begin{align}
d_1^{LB}(u,p) = d_1(u,p) + \frac{\text{Processor Load}}{\text{Load Factor}}
\label{equ:lb_embed}
\end{align}

The embed routing has all the benefits of smart routing. This routing scheme divides
the active regions (based on query workloads)
of the graph into $P$ partitions in an overlapping manner, and assigns them to the processors' cache. Moreover,
it dynamically adapts the partitions with new workloads. Therefore, {\em
we are able to bypass the expensive graph partitioning and re-partitioning problems to the existing cache replacement policy of the
query processors}. This shows the effectiveness of embed routing.

\spara{Dealing with Graph Updates.}
Due to pre-assignment of node co-ordinates, embed routing is less flexible with respect to graph updates. When a new node is added,
we compute its distance from the landmarks, and then assign co-ordinates to the node by applying the Simplex Downhill algorithm. Edge
updates and node deletions are handled in a similar method as discussed for landmark routing. We recompute the entire preprocessing
step periodically in an off-line manner to deal with a significant number of graph updates.

%% file: experiments.tex
\vspace{-3mm}
\section{EXPERIMENTAL RESULTS}
\label{sec:experiments}
\vspace{-1mm}
\subsection{Environment Setup}
\vspace{-2mm}
\spara{$\bullet$ Cluster Configuration.} We perform experiments on a cluster of 12 servers having 2.4 GHz
{\sf Intel} {\sf Xeon} processors, and interconnected by 40 Gbps Infiniband, and also by
10 Gbps Ethernet. Most experiments use a single core of each server with the following
configuration: 1 server as router, 7 servers in the processing tier, 4 servers
in the storage tier; and communication over Infiniband with remote direct memory
access ({\sf RDMA}). Infiniband is a modern and  mainstream network technology \cite{LPEK15,BCGKZ16},
that achieves low latency and high throughput, e.g., infiniband allows {\sf RDMA} in a
few microseconds. We use a limited main memory (0$\sim$4GB)
as the cache of processors. Our codes are implemented in C++.

To implement our storage tier, we use {\sf RAMCloud} \cite{OAEK10},
which provides high throughput and very low read/write latency, in
the order of 5-10 $\mu$s for every put/ get operation. It is able to achieve this efficiency
because it keeps all stored values in memory as a distributed key-value store,
where a key is hashed to determine on which server the corresponding key-value pair will be stored.
{\sf RAMCloud} also provides high memory utilization because of its log-structured design,
and ``continuous availability'' by harnessing large scale to recover from system failure in a few seconds.

The graph is stored as an adjacency list --- every node-id
in the graph is the key, and the corresponding value is an array of its 1-hop neighbors.
The graph is partitioned across storage servers via {\sf RAMCloud}'s
default and inexpensive hash partitioning scheme, {\sf MurmurHash3}
over graph nodes.

\spara{$\bullet$ Datasets.} We summarize our data sets in Table~\ref{tab:data}. As explained in Section~\ref{sec:system},
we store both in- and out-neighbors. 

\noindent {\bf WebGraph:} The uk-2007-05 web graph (http://law.di. unimi.it/ datasets.php)
is a collection of UK web pages, which are represented as nodes. A hyperlink in page $u$ to page $v$
is denoted by an edge between them.
\begin{table} [tb!]
\vspace{-3mm}
\scriptsize
\vspace{-2mm}
\centering
\begin{tabular} { l|ccc }
{\textsf{Dataset}} & {\textsf{\# Nodes}}  & {\textsf{\# Edges}}   &   {\textsf{Size on Disk (Adj. List File)}} \\ \hline\hline
 {\em WebGraph}    &     105\,896\,555    &  3\,738\,733\,648     &             60.3 GB      \\ \hline
 {\em Friendster}  &     65\,608\,366     &  1\,806\,067\,135     &             33.5 GB      \\ \hline
 {\em Memetracker} &     96\,608\,034    &   418\,237\,269        &             8.2 GB       \\ \hline
 {\em Freebase}    &     49\,731\,389    &  46\,708\,421         &              1.3 GB       \\ \hline
  \end{tabular}
\caption{\small Graph datasets}
\label{tab:data}
\vspace{-11mm}
\end{table}
\begin{figure*}[tb!]
\centering
\subfigure [\small {\em WebGraph}]{
\includegraphics[scale=0.3]{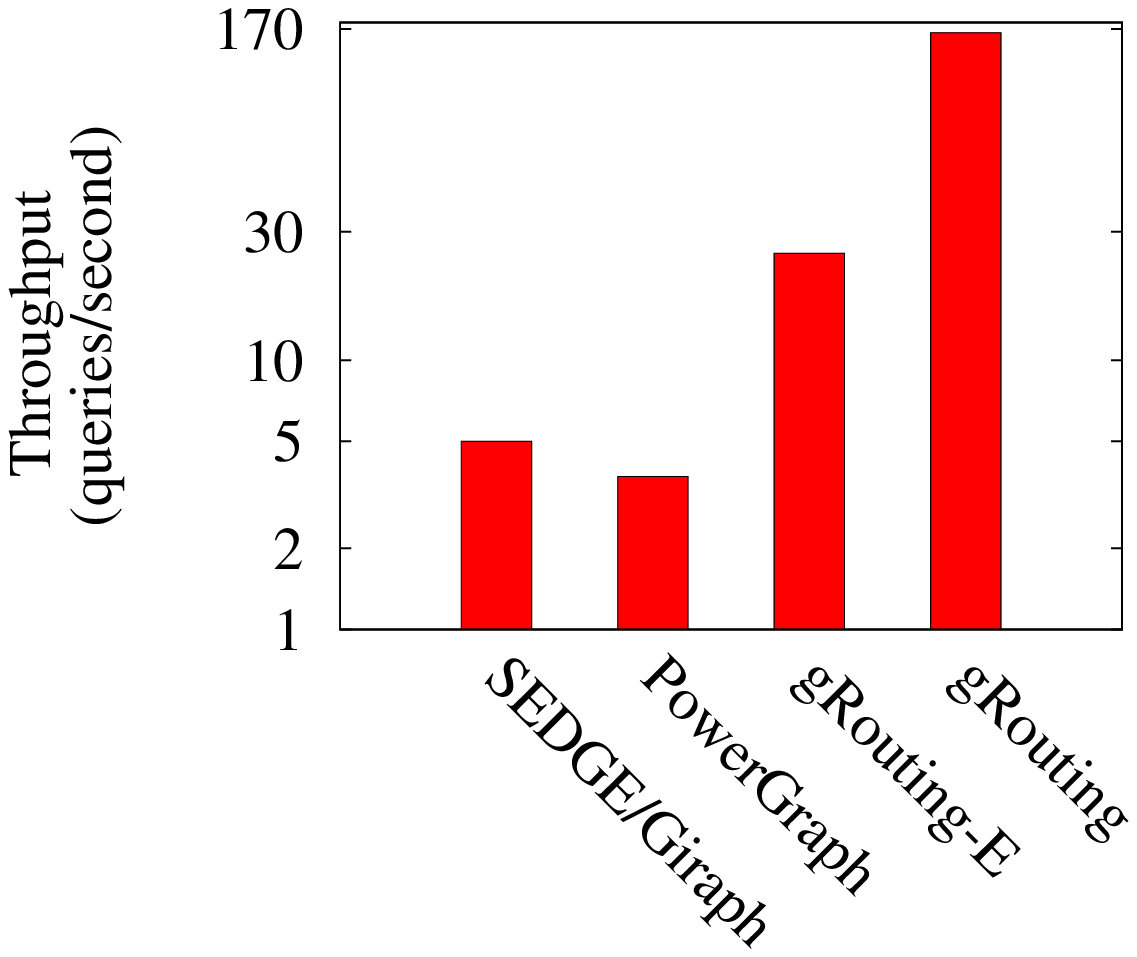}
\vspace{-2mm}
\label{fig:compare_web}
}
\subfigure [\small {\em MemeTracker}] {
\includegraphics[scale=0.3]{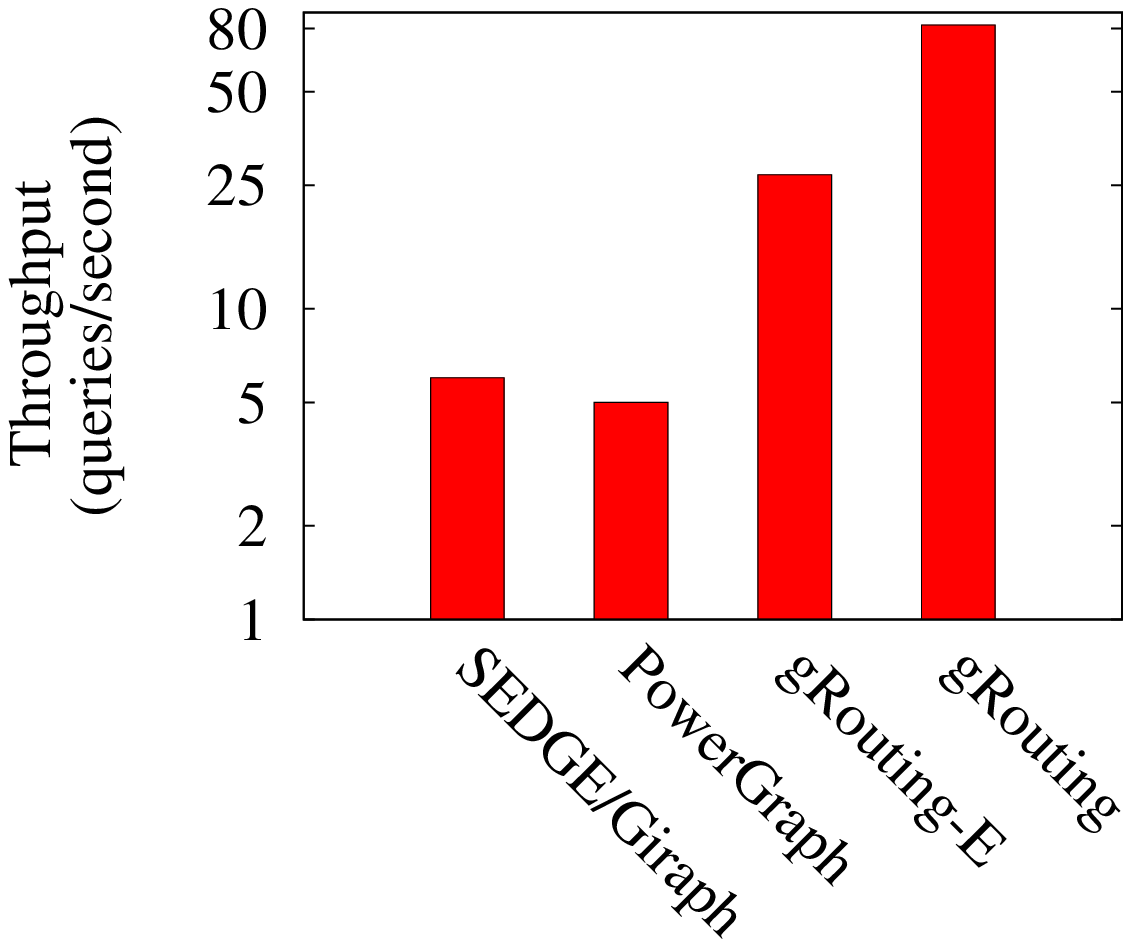}
\vspace{-2mm}
\label{fig:compare_meme}
}
\subfigure [\small {\em Freebase}] {
\includegraphics[scale=0.3]{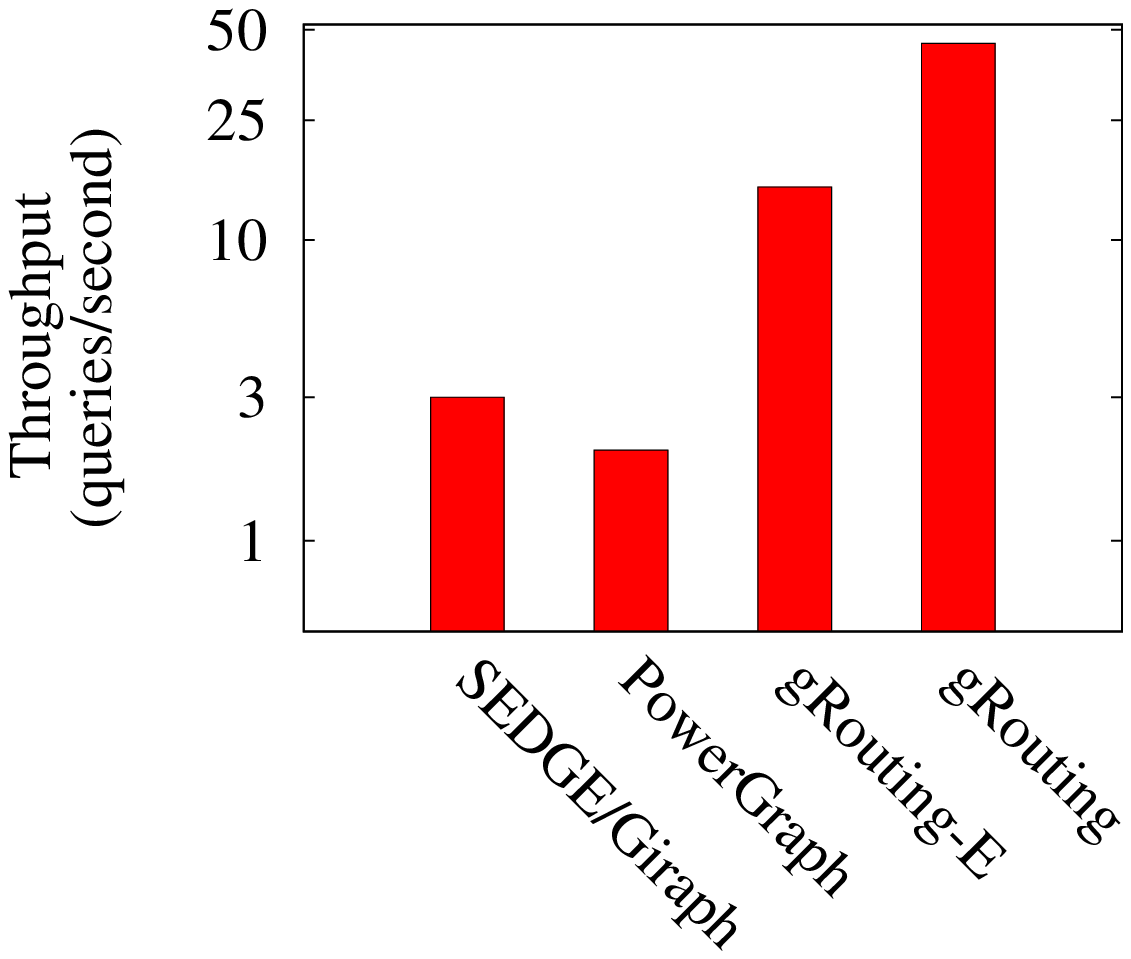}
\vspace{-2mm}
\label{fig:compare_freebase}
}
\vspace{-4mm}
\caption{\small Throughput comparison
}
\label{fig:comparison}
\vspace{-5mm}
\end{figure*}
\begin{figure*}[tb!]
\centering
\subfigure [\small Throughput]{
\includegraphics[scale=0.3]{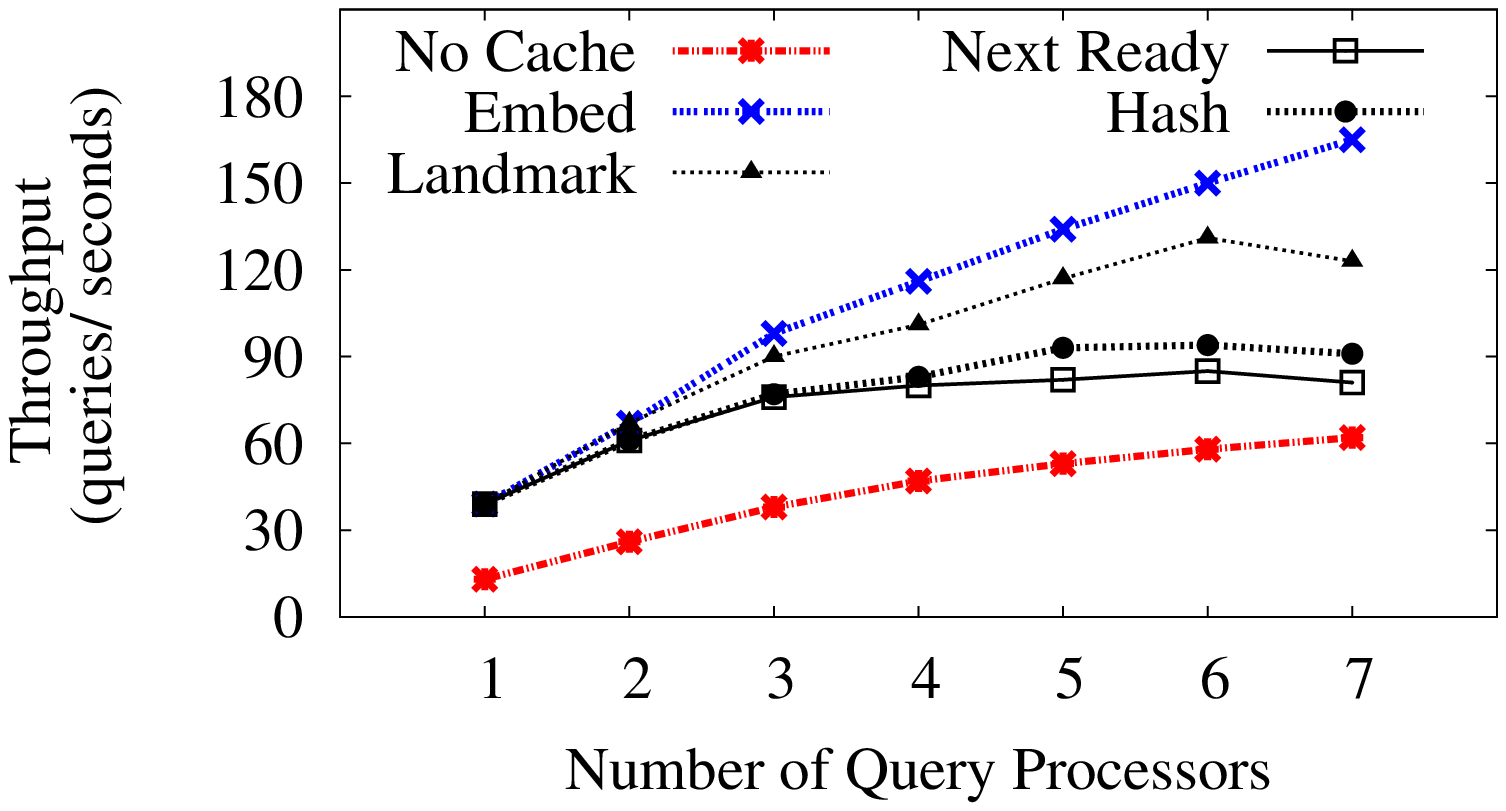}
\vspace{-2mm}
\label{fig:scale_throughput}
}
\subfigure [\small Cache Hits] {
\includegraphics[scale=0.3]{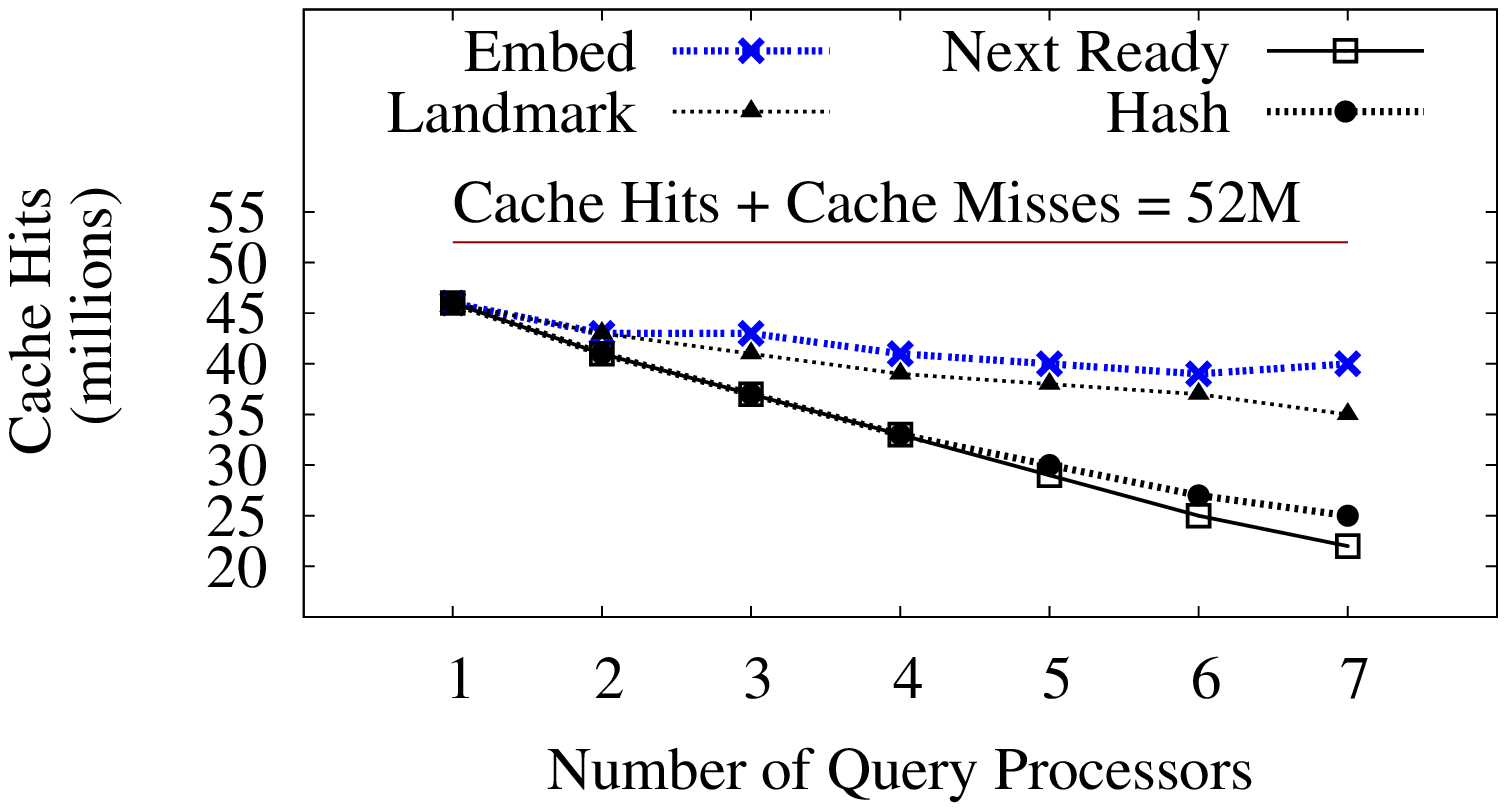}
\vspace{-2mm}
\label{fig:scale_hits}
}
\subfigure [\small Throughput] {
\includegraphics[scale=0.3]{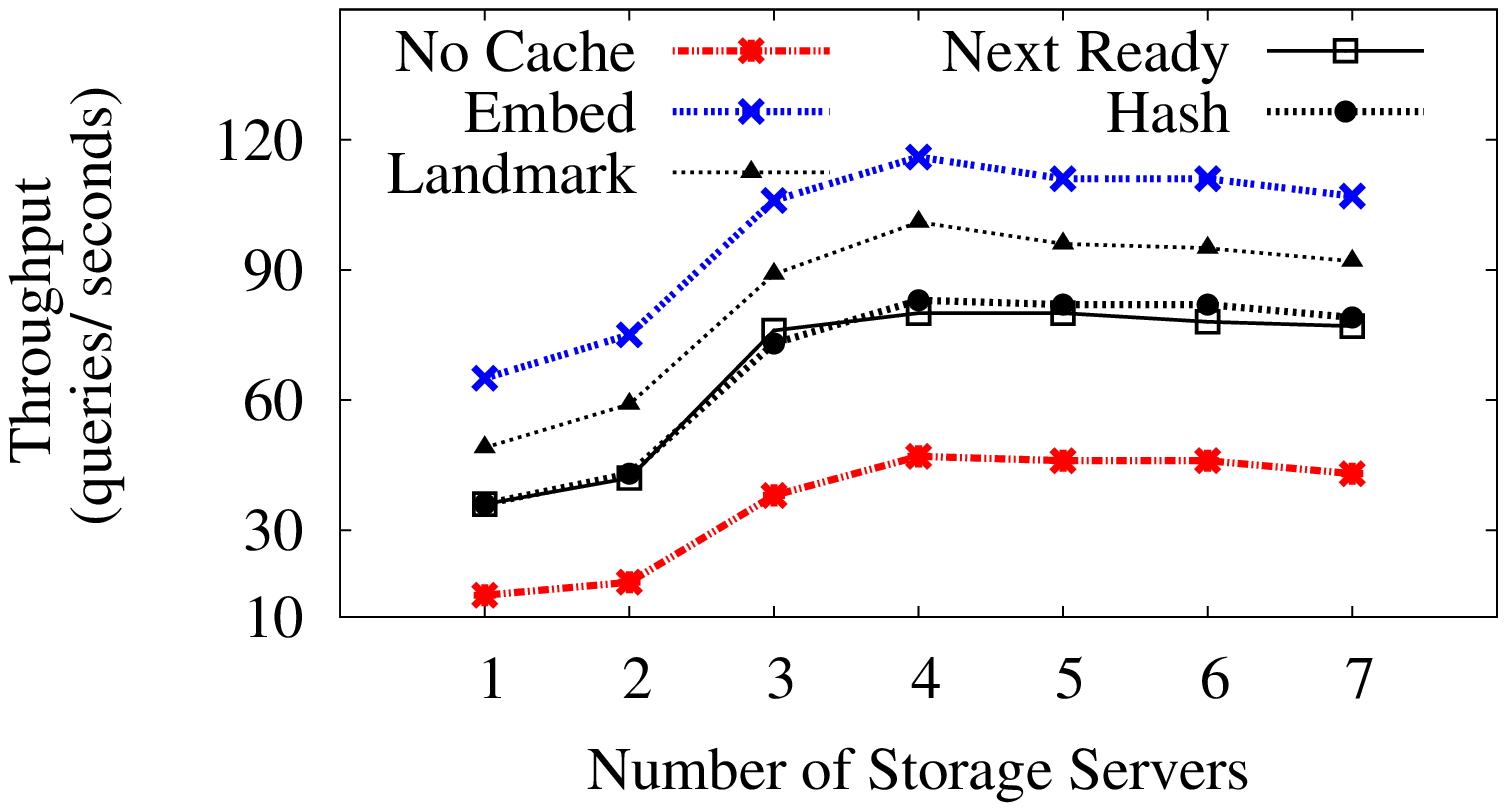}
\vspace{-2mm}
\label{fig:scale_storage}
}
\vspace{-4mm}
\caption{\small Performance with varying number of query processors and storage servers, {\em WebGraph}}
\label{fig:scalability_flex_qp}
\vspace{-5mm}
\end{figure*}
\begin{figure*}[t!]
\centering
\subfigure [\small Response time]{
\includegraphics[scale=0.3]{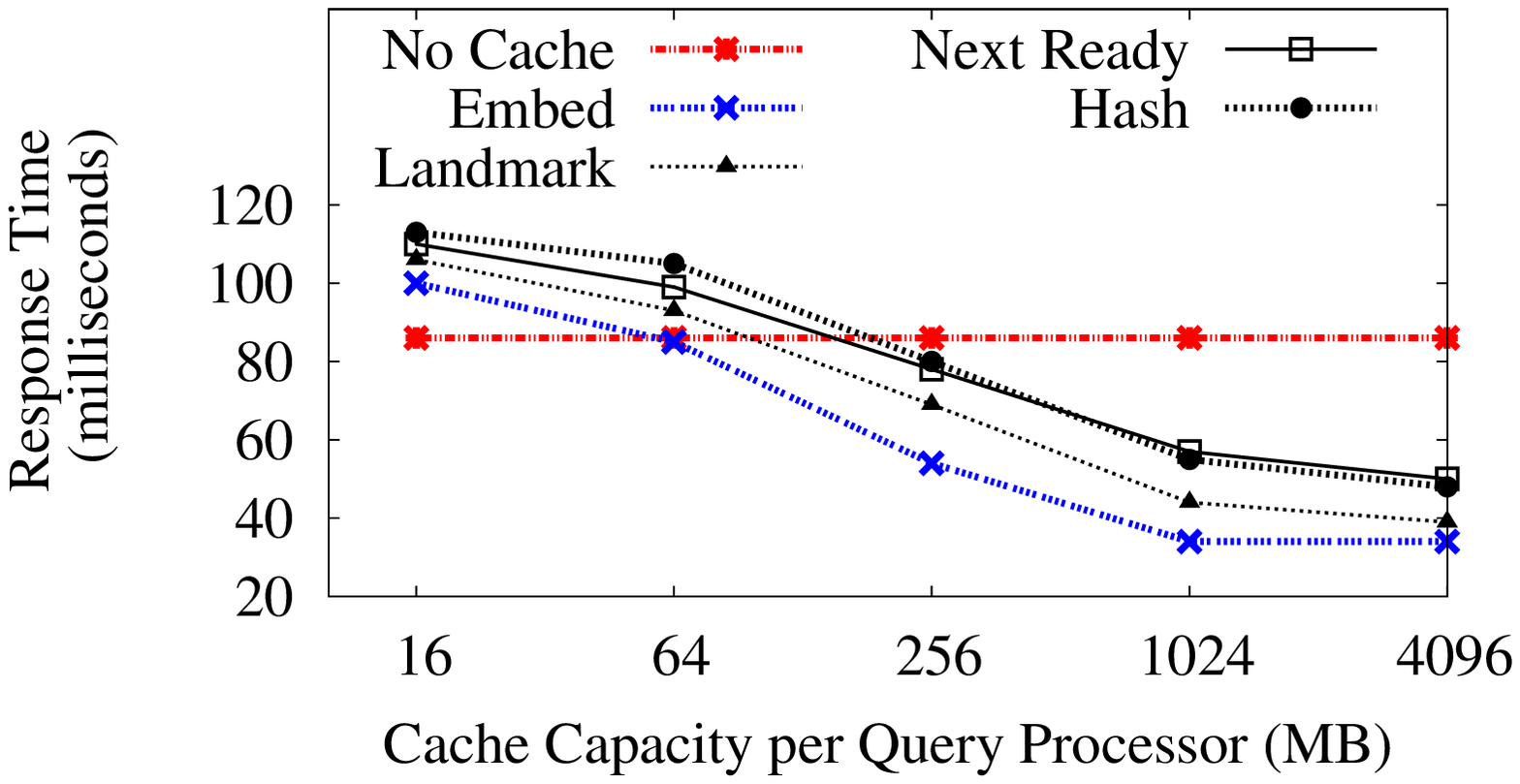}
\vspace{-2mm}
\label{fig:cache_response}
}
\subfigure [\small Cache hits] {
\includegraphics[scale=0.3]{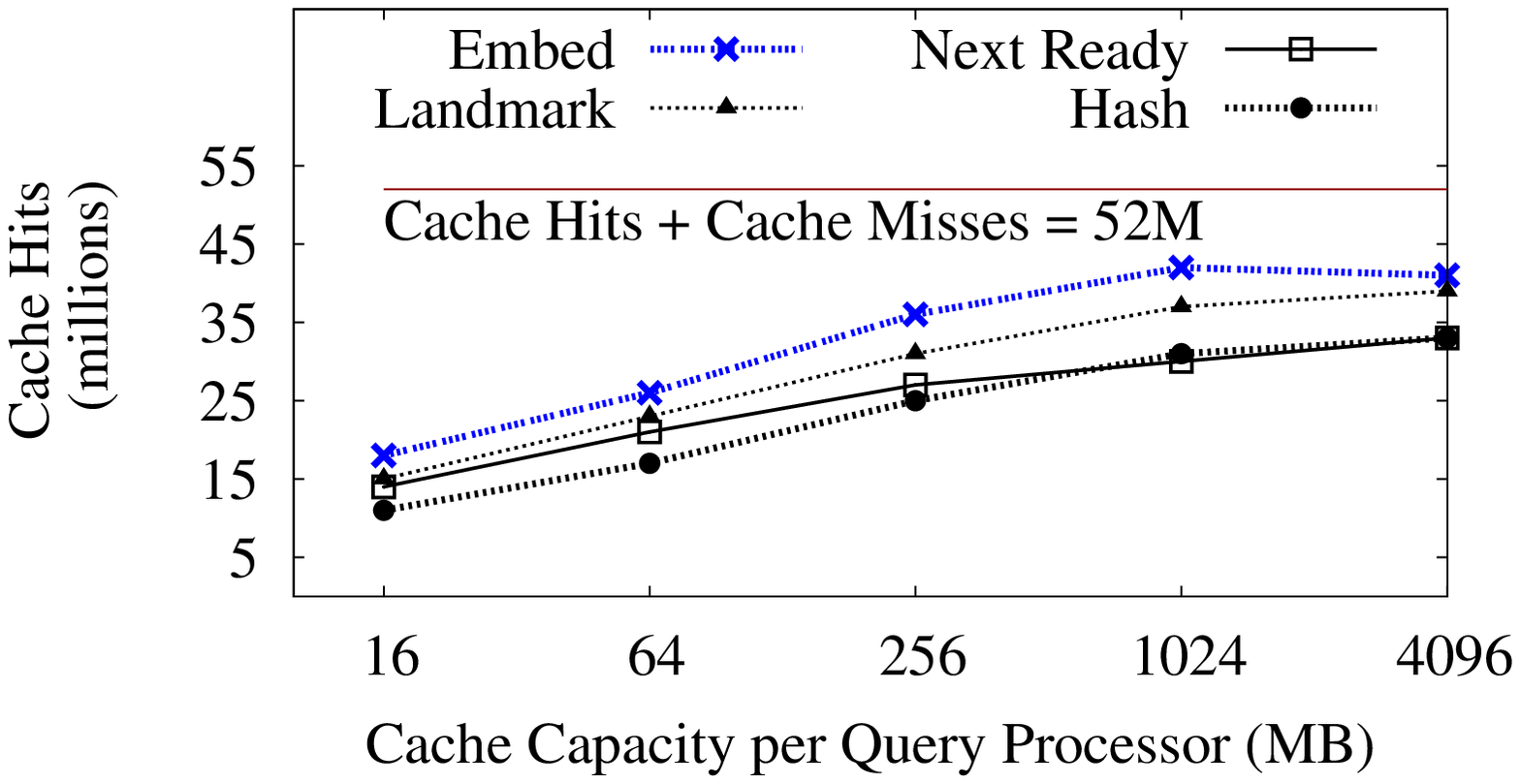}
\vspace{-2mm}
\label{fig:cache_hits}
}
\subfigure [\small Min. cache required to reach \newline  \hspace*{1em} No-Cache's response time] {
\includegraphics[scale=0.3]{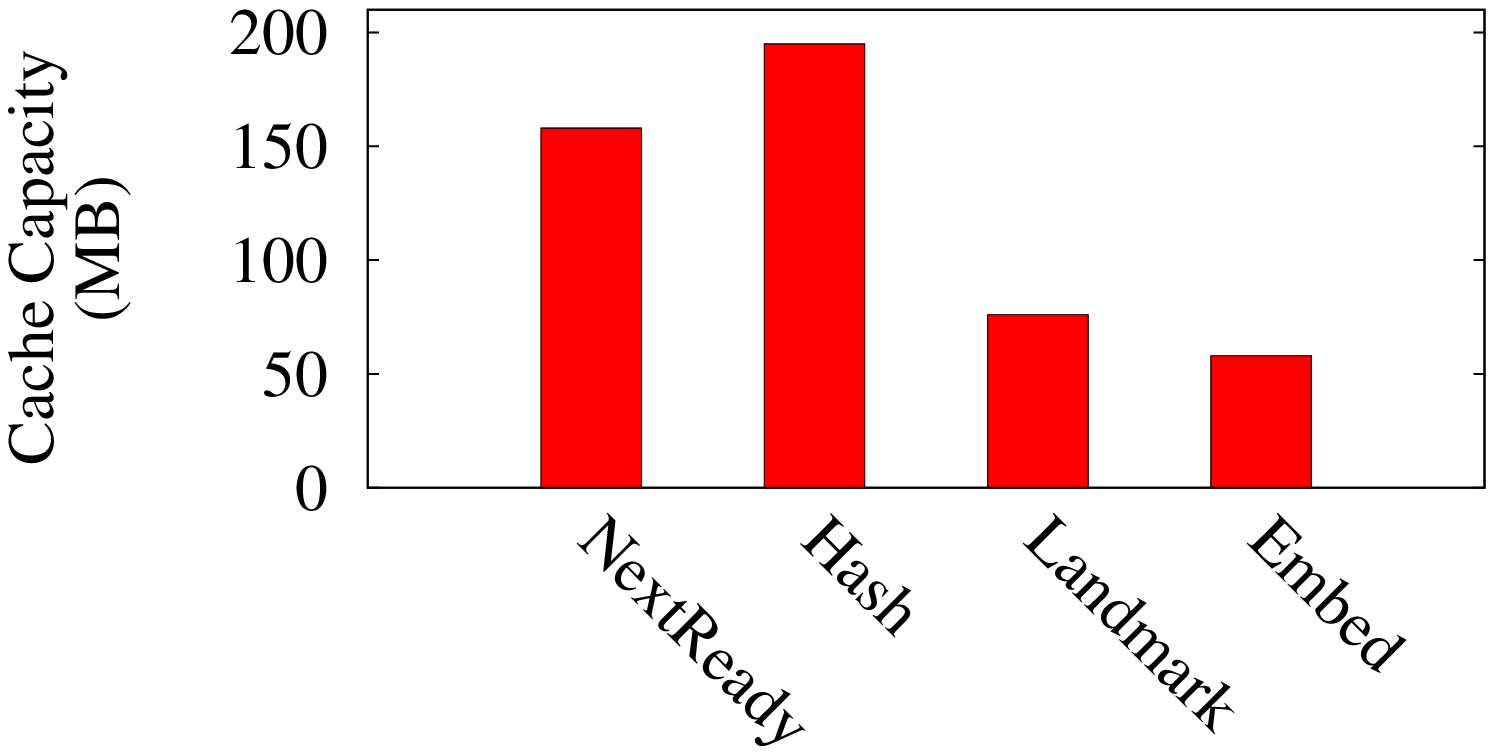}
\vspace{-2mm}
\label{fig:memory}
}
\vspace{-5mm}
\caption{\small Impact of cache size, {\em WebGraph}}
\label{fig:cache}
\vspace{-5mm}
\end{figure*}

\noindent {\bf Friendster:} We download this graph from snap.stanford.edu, which is a subgraph of the
original {\em Friendster} social network. The nodes represent users and the edges denote friendship links.

\noindent {\bf Memetracker:} This dataset (snap.stanford.edu) tracks quotes and phrases that appeared frequently from
August 1 to October 31, 2008 across online news spectrum. We consider the documents as nodes and
hyper-links as edges.

\noindent {\bf Freebase:} We download the {\em Freebase} knowledge graph from http://www.freebase.com/.
Nodes are named entities (e.g., Google) or abstract concepts (e.g., Asian people), and edges denote
relations (e.g., founder) between two nodes.

\spara{$\bullet$ Online Query Workloads.}  We consider three online graph queries \cite{YYZK12}, all require traversals up to $h$ hops: (1) {\em $h$-hop Neighbor Aggregation}: Count the
number of $h$-hop neighbors of a query node. (2) {\em $h$-step Random Walk with Restart}: The query starts at a node, and at each successive step
jumps to one of its neighbors with equal probability, or returns to the query node with a small probability. The query consists of $h$ steps.
(3) {\em $h$-hop Reachability}: Given a source and a target node, find if the target is reachable from the source within $h$-hops. In our experiments, we
consider a uniform mixture of above queries.
We simulate a scenario when queries are drawn from a hotspot region;
and the hotspots change over time. In particular, we select 100 nodes from the graph uniformly at random.
Then, for each of these nodes, we select 10 different query nodes
which are at most $r$-hops away from that node. Thus, we generate 1000 queries; every 10 of them are from one hotspot region, and the pairwise distance between any two nodes from the same hotspot is
at most $2r$. Finally, all queries from the same hotspot are grouped together and sent consecutively. We report our
results averaged over 1000 queries.

In order to realize the effect of topology-aware locality, we consider smaller values of $r$ and $h$, e.g., $r=2$
and $h=2$ in most of our experiments. We also analyze the impact of varying $h (=1\sim 3)$ and $r (=1\sim 2)$
in Section~\ref{sec:exp_q_parameter}.

\spara{$\bullet$ Evaluation Metrics.}  We measure performance as follows.


\noindent{\bf Query Response Time} measures the average time required to answer one query.


\noindent{\bf Query Processing Throughput} measures the number of queries that can be processed per unit time.


\noindent{\bf Cache Hit Rate:} We report cache hit rates, since higher cache hit rates reduce the query response time.
Consider $t$ queries $q_1, q_2,$ $\ldots,$ $q_t$ received successively by the router. For simplicity, let us assume that each query
retrieves all $h$-hop neighbors of that query node (i.e., $h$-hop neighborhood aggregation).
We denote by $|N_h(q_i)|$ the number of nodes within $h$-hops from $q_i$. Among them, we assume
that $|N^c_h(q_i)|$ number of nodes are found in the query processors' cache. We define cache hit and miss rates
as:
\vspace{-2mm}
\begin{align}
&\displaystyle \text{Cache Hit Rates} := \sum_{i=1}^t |N^c_h(q_i)| & \\
&\displaystyle \text{Cache Miss Rates} := \sum_{i=1}^t \left(|N_h(q_i)| - |N^c_h(q_i)|\right) &
\end{align}

\vspace{-2mm}
\spara{$\bullet$ Parameter Setting.} We find that {\em embed routing performs the best compared to three other routing strategies}.
We also set the following parameter values since they perform
the best in our implementation. We shall, however, demonstrate sensitivity of our routing algorithms with these parameters in Section~\ref{sec:sensitivity}.

We use maximum 4GB cache in each query processor. All experiments are performed with the cache initially empty (cold cache). The number of landmarks $|L|$ is set as
96 with at least 3 hops of separation from each other. For graph embedding, 10 dimensions are used. Load Factor (which impacts query stealing) is set as 20,
and the smoothing parameter $\alpha = $ 0.5.

In order to realize how our routing schemes perform compared to the scenario when there is no cache in query processors, we consider an additional
``no-cache'' scheme. In this mode, all queries are routed following the next ready technique; however, as there is no cache in query processors,
there will be no overhead due to cache lookup and maintenance.

\spara{$\bullet$ Compared Systems.}
To the best of our knowledge, smart query routing logic for distributed graph querying
is novel.
Moreover, decoupled architecture and our smart routing logic,
being generic, can benefit many graph querying systems.
Nevertheless, we compare {\sf gRouting} with two distributed
graph processing systems: {\sf SEDGE/Giraph} \cite{YYZK12} and {\sf PowerGraph} \cite{GLGBG12}.
Other recent graph querying systems, e.g., \cite{SEHM13,NFGKLLMPPSSV13}
are not publicly available for a direct comparison.

{\sf SEDGE} \cite{YYZK12} was developed for $h$-hop traversal queries
(i.e., random walk, reachability, and neighborhood aggregation), on top of {\sf Giraph} or Google's {\sf Pregel}
system \cite{MABDHLC10}. It follows in-memory, vertex-centric, bulk-synchronous parallel model.
{\sf SEDGE} employs {\sf ParMETIS} software \cite{KK98} for graph partitioning and re-partitioning.
Their re-partitioning method requires prior knowledge about future queries,
whereas the proposed {\sf gRouting} scheme is adaptive with workload changes, and it does not require any
prior knowledge of future queries. Even so, we compare against {\sf SEDGE} with re-partitioning,
since {\sf SEDGE} provides the best throughput with re-partitioned graphs.

{\sf PowerGraph} \cite{GLGBG12} follows in-memory, vertex-centric, asynchronous
gather-apply-scatter model. For efficient implementation of our $h$-hop traversal queries, we ensure that {\em only} the required
nodes are active at any point of time, that is, in the beginning, only the query node is active, and
each active node then activates its neighbors, until all the $h$-hop neighbors from the query
nodes are activated. {\sf PowerGraph} also employs a sophisticated node-cut based graph partitioning method,
for improved query throughput.
\vspace{-3mm}
\subsection{Comparison with Distributed \\ Graph Systems}
%
In Figure~\ref{fig:comparison}, we compare {\sf gRouting} with two distributed
graph processing systems, {\sf SEDGE/Giraph} \cite{YYZK12} and {\sf PowerGraph} \cite{GLGBG12}.
As these systems run on Ethernet, we consider a version of {\sf gRouting}
on Ethernet ({\sf gRouting-E}). We consider 12 machines configuration of {\sf SEDGE}
and {\sf PowerGraph}, since query processing and graph storage in them
are coupled on same machines. On the contrary, we fix the number of routing, processing, and storage servers as 1, 7 and 4, respectively.
The average 2-hop neighborhood size varies from 10K$\sim$60K nodes over our datasets.

We find that our throughput, with hash partitioning, and over Ethernet, is 5$\sim$10 times better than
two existing systems ({\sf SEDGE} and {\sf PowerGraph}) with expensive graph partitioning and re- partitioning
strategies. The re-partitioning in {\sf SEDGE} requires around $1$ hour and also apriori information on
future queries, whereas {\sf PowerGraph} graph partitioning finishes in $30$ min. On the contrary, {\sf gRouting}
performs lightweight hash partitioning over graph nodes, and does not require any prior knowledge of the future
workloads. Moreover, our throughput over Infiniband is 10$\sim$35 higher than these systems.
{\em These results show the usefulness of smart query routing over expensive graph partitioning and re-partitioning schemes}.

In the following, we report scalability results, impact of cache sizes, and graph updates
over {\em Webgraph} due to brevity, and also because this is the largest dataset in our experiments.
We shall, however, present empirical results over other datasets in Section~\ref{sec:exp_datasets}.
\vspace{-6mm}
\subsection{Scalability and Deployment Flexibility}
%
One of the main benefits of separating processing and storage tiers is deployment flexibility --- they can be scaled-up independently,
which we investigate in the following.

\vspace{-1mm}
\spara{Processing Tier:} We vary the number of processing servers from 1 to 7, while using 1 router and 4 storage servers.
In Figure~\ref{fig:scale_throughput}, we show throughput with varying number of processing servers.
Corresponding cache hit rates are presented in Figure~\ref{fig:scale_hits}.
For these experiments, we assume that each query processor has sufficient cache capacity (4GB) to store
the results of all 1000 queries (i.e, adjacency lists of 52M nodes, shown in Figure~\ref{fig:scale_hits}).
Since, for every experiment, we start with an empty cache, and then send the
same 1000 queries in order, maximum cache hit happens when there is only one query processor. As we increase
the number of query processors, these queries get distributed and processed by different processors,
thus cache hit rate generally decreases. This is more evident for our baseline routing schemes, and we find that their
throughput saturates with 3$\sim$5 servers. These findings demonstrate the usefulness
of smart query routing: {\em To maintain same cache hit rate, queries must
be routed intelligently. Since Embed routing is able to sustain almost
same cache hit rate with many query processors (Figure~\ref{fig:scale_hits}),
its throughput scales linearly with query processors}.

\vspace{-1mm}
\spara{Storage Tier:} We next vary the number of storage servers from 1 to 7, whereas 1 server is used as the router
and 4 servers as query processors (Figure~\ref{fig:scale_storage}).
When we use 1 storage server, we can still load the entire 60GB {\em Webgraph} on the main memory of that server,
since each of our servers has sufficient {\sf RAM}.
The throughput is the least when there is only one storage server.
We observe that 1$\sim$2 storage servers are insufficient to handle the demand created by
4 query processors. However, with 4 storage servers, the throughput saturates, since the bottleneck is transferred to query processors.
This is evident from our previous results --- the throughput with 4 query processors was about 120 queries per second (Figure~\ref{fig:scale_throughput}),
which is the same throughput achieved with 4 storage servers in the current experiments.
\vspace{-2mm}
\subsection{Impact of Cache Sizes}
In previous experiments, we assign 4GB cache to each processor, which was large enough for our queries; and we never discarded anything from the cache.
We next perform experiments when it needs to evict cache entries. In Figure~\ref{fig:cache}, we present average response times with various cache capacities.
At the largest, with 4GB cache per processor, no eviction occurs. Therefore, there is no additional performance gain by increasing the cache capacity. On the other extreme,
having cache with less than 64MB per processor results in worse response times than what was obtained with no-cache scheme, represented
by the horizontal red line (86ms in Figure~\ref{fig:cache}). When the cache does not have much space, it ends up evicting entries that might have been useful
in the future. Hence, there are not enough cache hits to justify its maintenance and lookup costs when the cache size is smaller than 64MB
per processor.

We also evaluate our routing strategies in terms of minimum cache requirement to achieve a response time of 86ms,
the break-even point of deciding whether or not to add a cache. Figure~\ref{fig:memory} shows
that smart routing schemes achieve this response time with a much lower cache requirement, as compared to that of the baselines.
These results illustrate that {\em our smart routings utilize the cache space well; and for the same amount of cache space, they
achieve lower response time compared to baseline routings}.
\vspace{-2mm}
\subsection{Preprocessing and Graph Updates}
\vspace{-2mm}
\spara{Preprocessing Time and Storage:} For landmarks routing, we compute the distance of every node to all landmarks, which can be evaluated by performing a
{\sf BFS} from each landmark. This takes about 35 sec for one landmark in {\em Webgraph} (Table~\ref{tab:preprocessing_time}),
and can be parallelized per landmark. For embed routing, in addition, we need to embed every node with respect to landmarks, which requires about 1 sec per node in {\em Webgraph}, and is again parallelizable per node.
\begin{table}[t!]
\begin{varwidth}[b]{0.5\linewidth}
\begin{tiny}
\begin{tabular} { c|cc }
{\textsf{Landmark}} & \multicolumn{2}{c}{\textsf{Embed}} \\
                            & embed/landmarks & embed/node        \\ \hline \hline
 35 sec                     &     36 sec      &  1 sec                \\ \hline
\end{tabular}
\caption{\small Preprocessing times}
\label{tab:preprocessing_time}
\vspace{-2mm}
\begin{tabular} { c|c||c }
{\textsf{Landmark}} & {\textsf{Embed}} & {\textsf Original graph size} \\ \hline \hline
 2.8 GB                     &     4GB                  &   60.3GB \\ \hline
 \end{tabular}
\caption{\small Preprocessing storage}
\label{tab:preprocessing_storage}
\end{tiny}
\vspace{-4mm}
\end{varwidth}
\hfill
\vspace{-3mm}
\begin{minipage}[b]{0.47\linewidth}
\centering
\vspace{-21mm}
\includegraphics[scale=0.25]{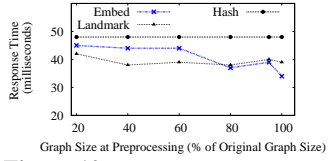}
\vspace{-8mm}
\captionof{figure}{\small Robustness with \newline graph updates}
\label{fig:update}
\vspace{-3mm}
\end{minipage}
\end{table}

The preprocessed landmark routing information consumes about 2.8GB storage space in case of {\em Webgraph}.
On the contrary, with embedding dimensionality 10, the {\em Webgraph} embedding size is only 4GB.
Both these preprocessed information are modest compared to the original {\em Webgraph} size, which is around 60GB (Table~\ref{tab:preprocessing_storage}).

\vspace{-1mm}
\spara{Graph Updates:}
In these experiments, we perform minimal changes (often none) to our preprocessed information with every graph update, and analyze how robust our routing schemes are with respect to
graph modification.
\begin{figure}[tb!]
\centering
\subfigure [\small Throughput]{
\includegraphics[scale=0.21]{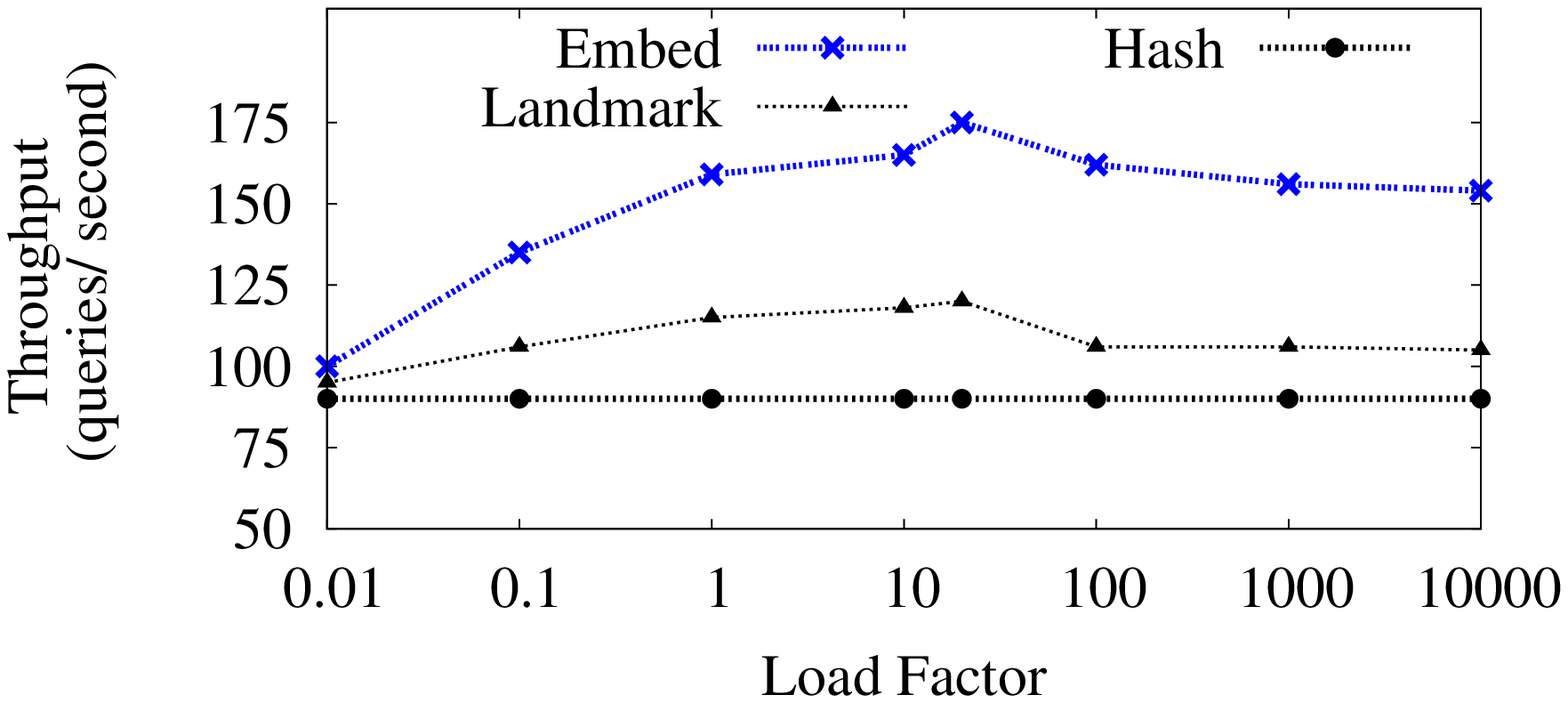}
\label{fig:load}
}
\subfigure [\small Response time] {
\includegraphics[scale=0.21]{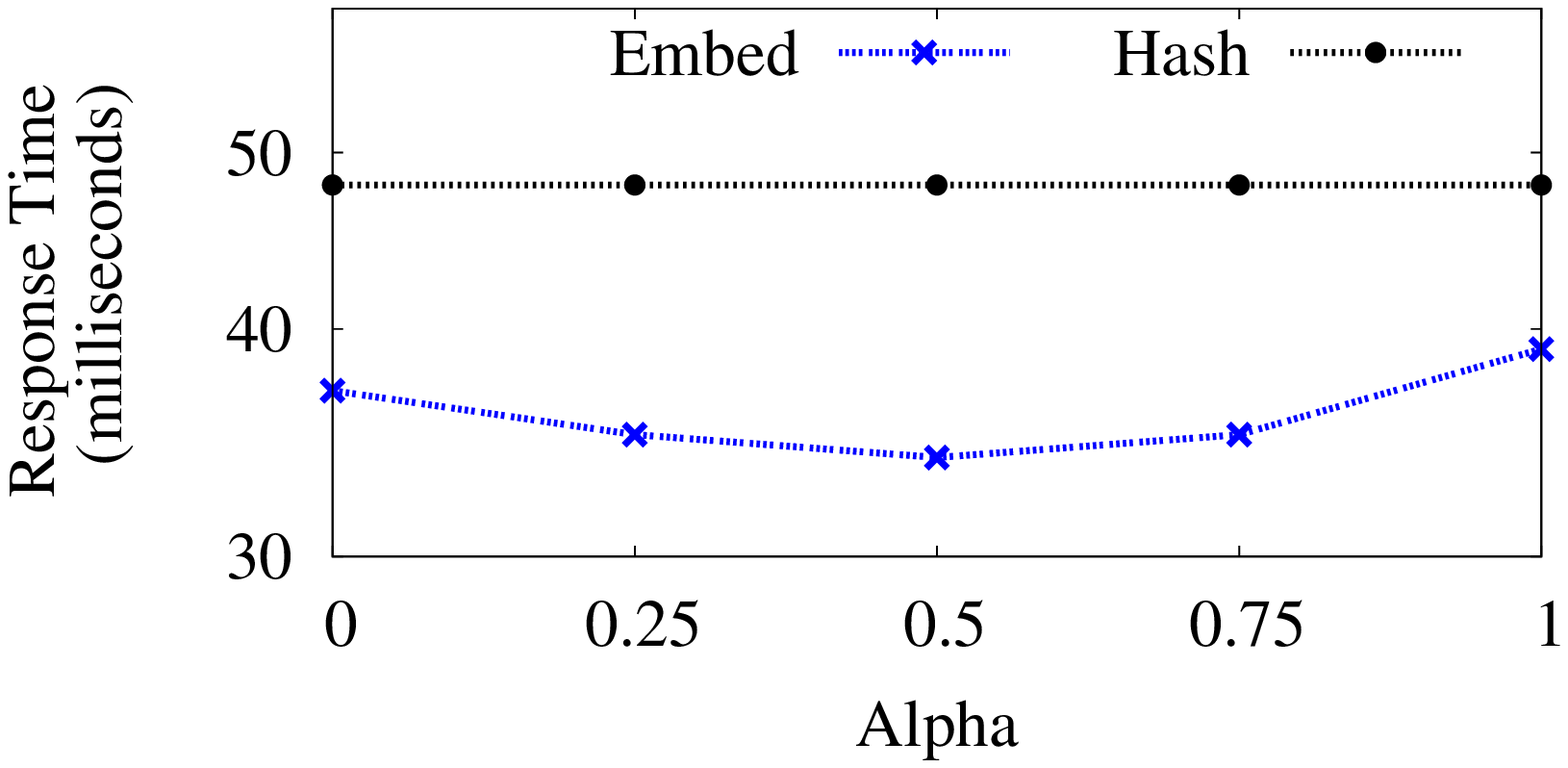}
\label{fig:alpha}
}
\vspace{-4mm}
\caption{\small Impact of load factor and smoothing parameter ($\alpha$)}
\label{fig:load_alpha}
\vspace{-5mm}
\end{figure}
\begin{figure}[tb!]
\centering
\subfigure [\small Relative error]{
\includegraphics[scale=0.23]{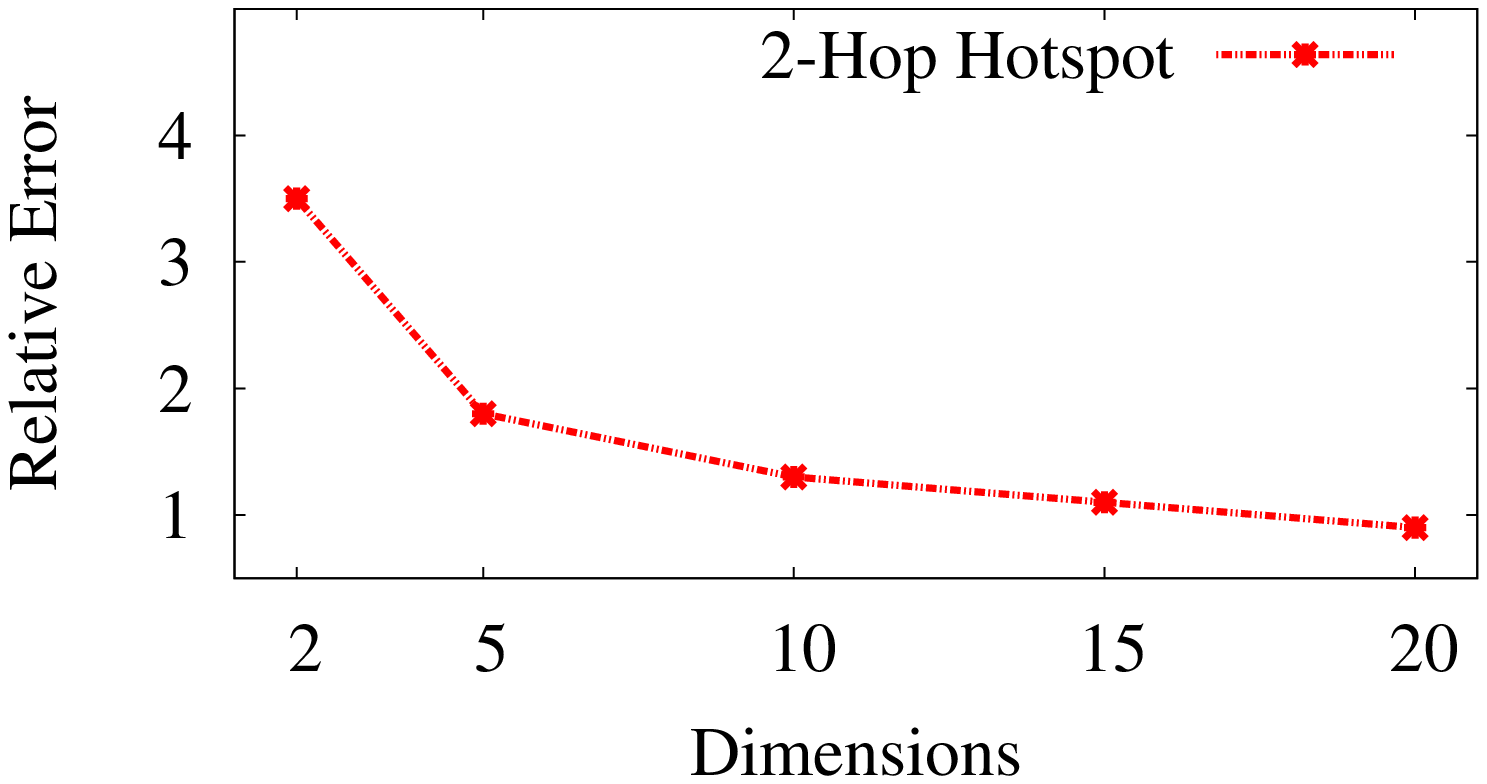}
\label{fig:dim_error}
}
\subfigure [\small Response time] {
\includegraphics[scale=0.23]{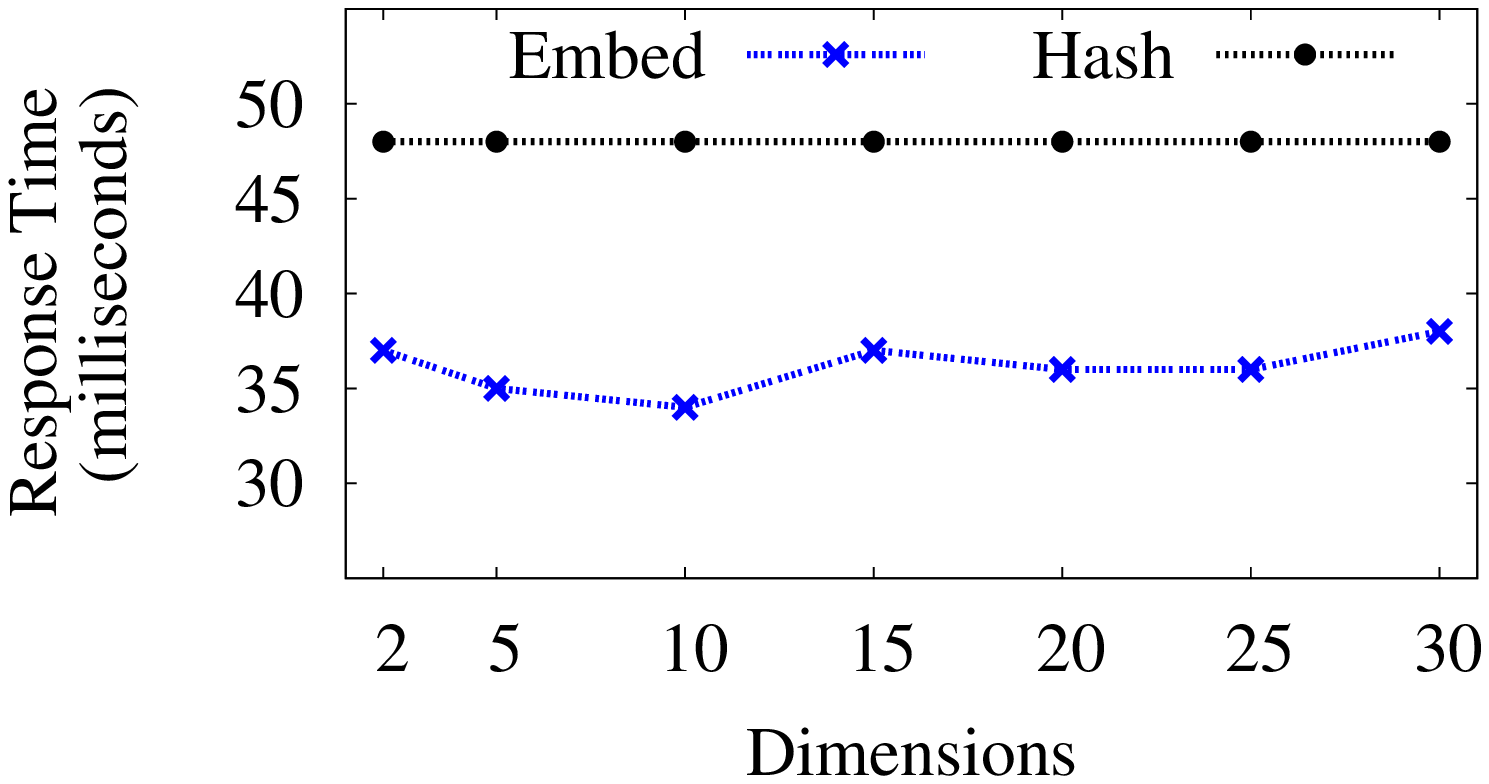}
\label{fig:dim_response}
}
\vspace{-4mm}
\caption{\small Impact of embedding dimensionality}
\label{fig:dimension}
\vspace{-5mm}
\end{figure}
\begin{figure}[tb!]
\centering
\subfigure [\small Response time]{
\includegraphics[scale=0.23]{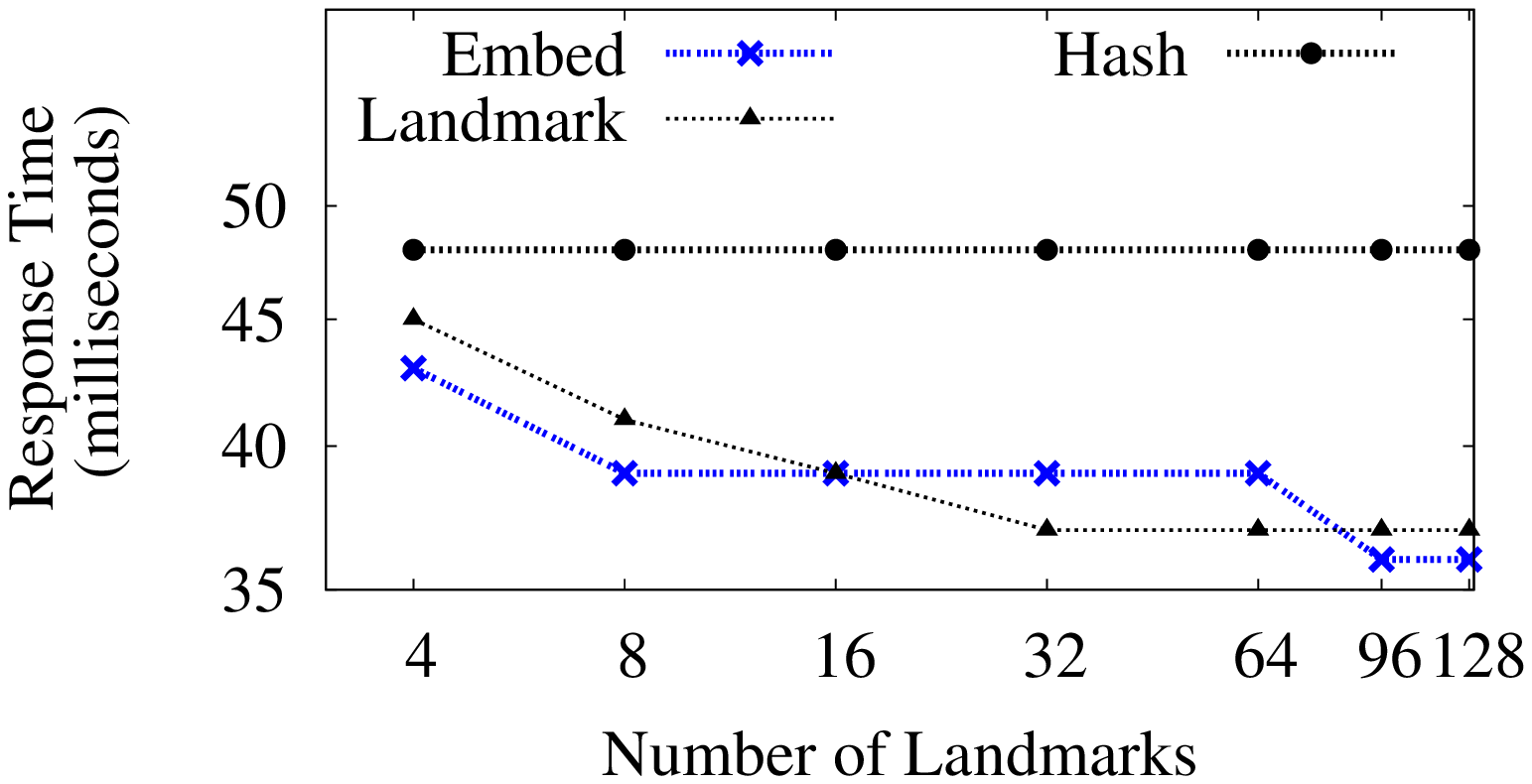}
\label{fig:no_landmark}
}
\subfigure [\small Response time] {
\includegraphics[scale=0.23]{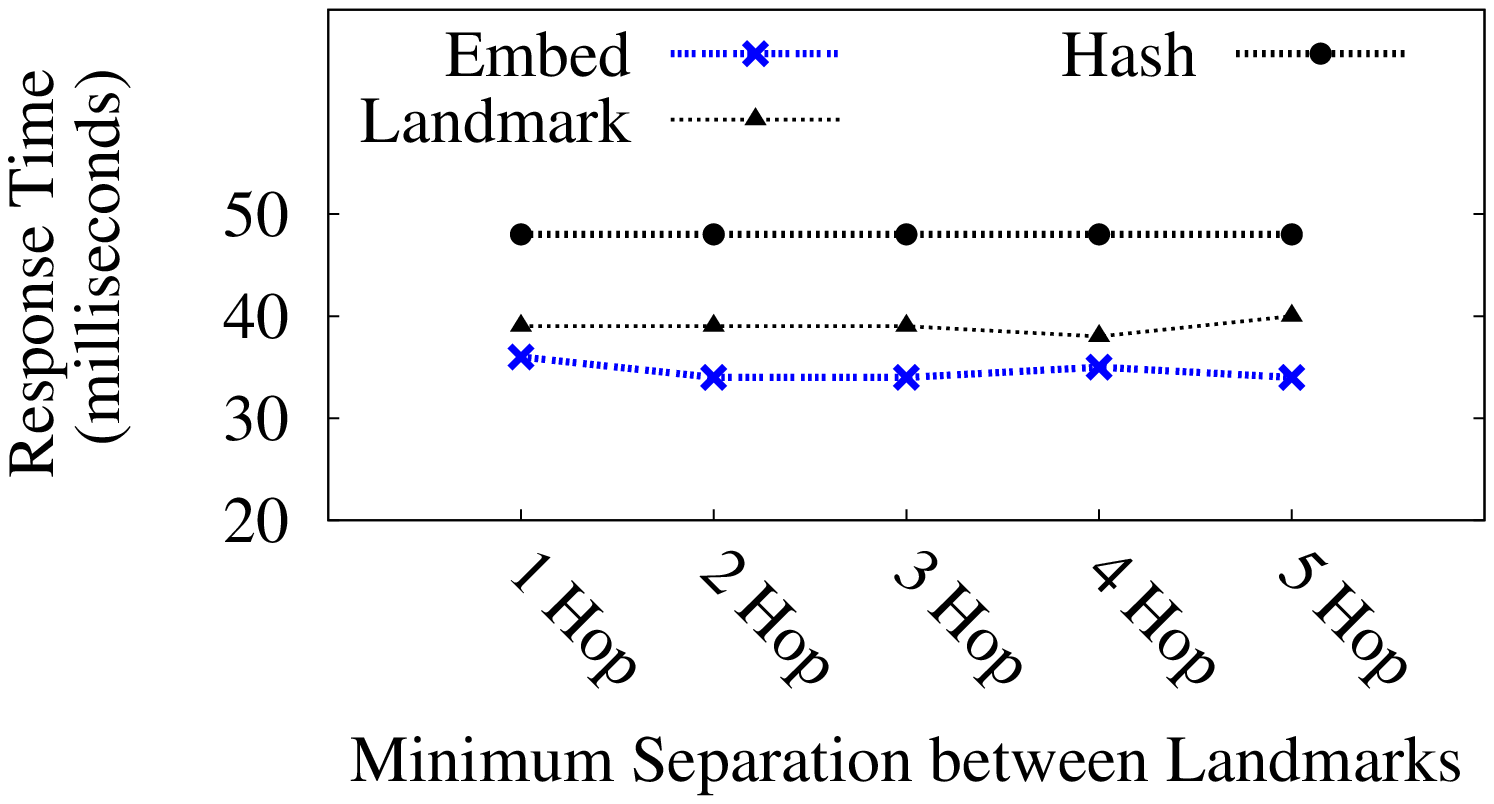}
\label{fig:separation_landmark}
}
\vspace{-4mm}
\caption{\small Impact of landmark numbers and their separation}
\label{fig:landmarks}
\vspace{-5mm}
\end{figure}

In particular, we preprocess a reduced subgraph of the original dataset.
For example, at 20\% of the original dataset (Figure~\ref{fig:update}), we select only 20\% of all nodes uniformly at random, and compute preprocessed information over the subgraph induced
by these selected nodes. However, we always run our query over the complete {\em Webgraph}. We incrementally compute the necessary information for the new nodes, as they
are being added, without changing anything on the preprocessed information of the earlier nodes. As an example, in case of embed routing, we only
compute the distance of a new node to the landmarks, and thereby find the coordinates of that new node. However, one may note that with the addition of every new node and its adjacent edges, the
preprocessed information becomes outdated (e.g., the distance between two earlier nodes might decrease). Since we do not change anything on the preprocessed
information, this experiment demonstrates the robustness of our method with respect to graph updates.

Figure~\ref{fig:update} depicts that {\em our smart routing schemes are robust for a small number of graph updates}. With embed routing, preprocessed information over the whole graph results in response time of 34 ms, whereas preprocessed information at 80\% of the graph results in response time of 37 ms (i.e., response time increases by only 3 ms). As expected, the response time deteriorates when preprocessing is performed on a smaller amount of graph data, e.g., with only 20\% graph data, response time increases to 44 ms, which is comparable to the response time of baseline hash routing (48 ms).
\begin{figure*}[tb!]
\centering
\subfigure [\small Response time]{
\includegraphics[scale=0.34]{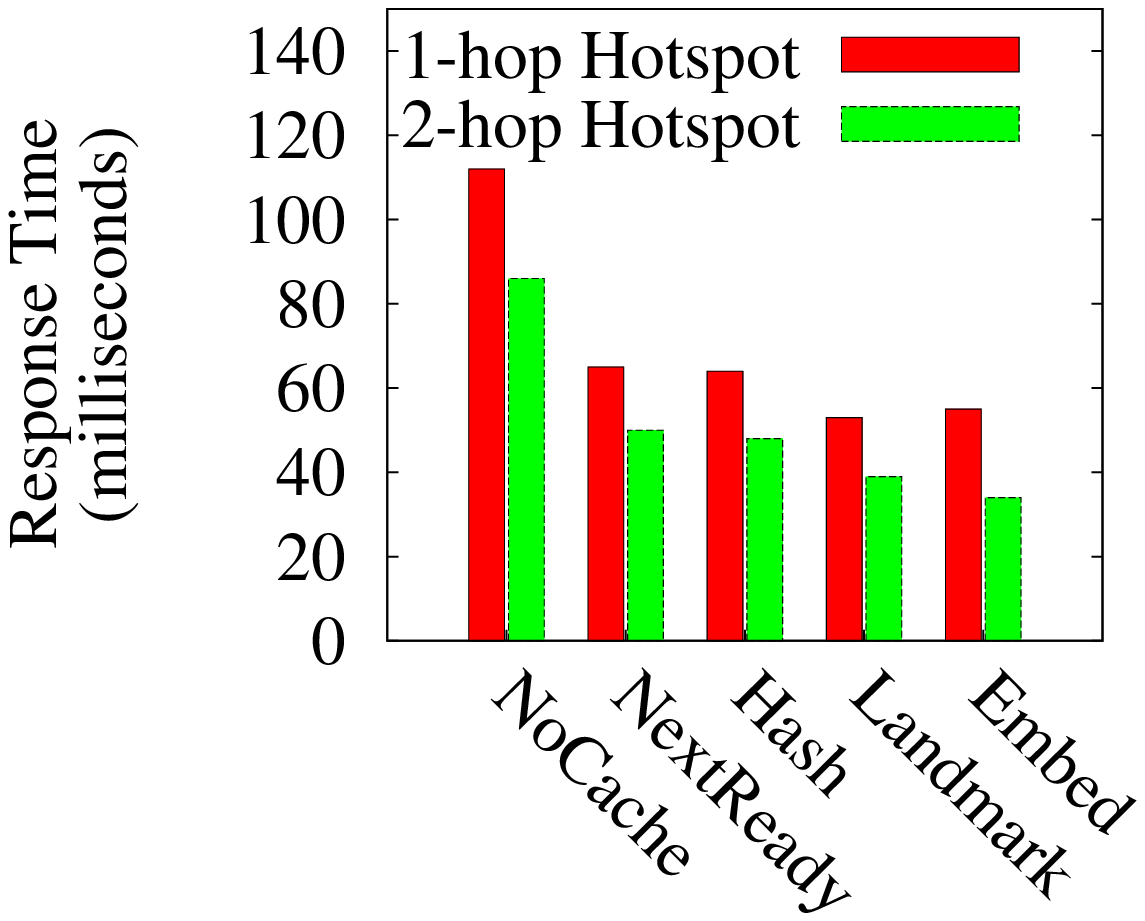}
\label{fig:response_time_web}
}
\subfigure [\small Cache hits and misses \newline  \hspace*{1.5em} (1-hop hotspot)] {
\includegraphics[scale=0.34]{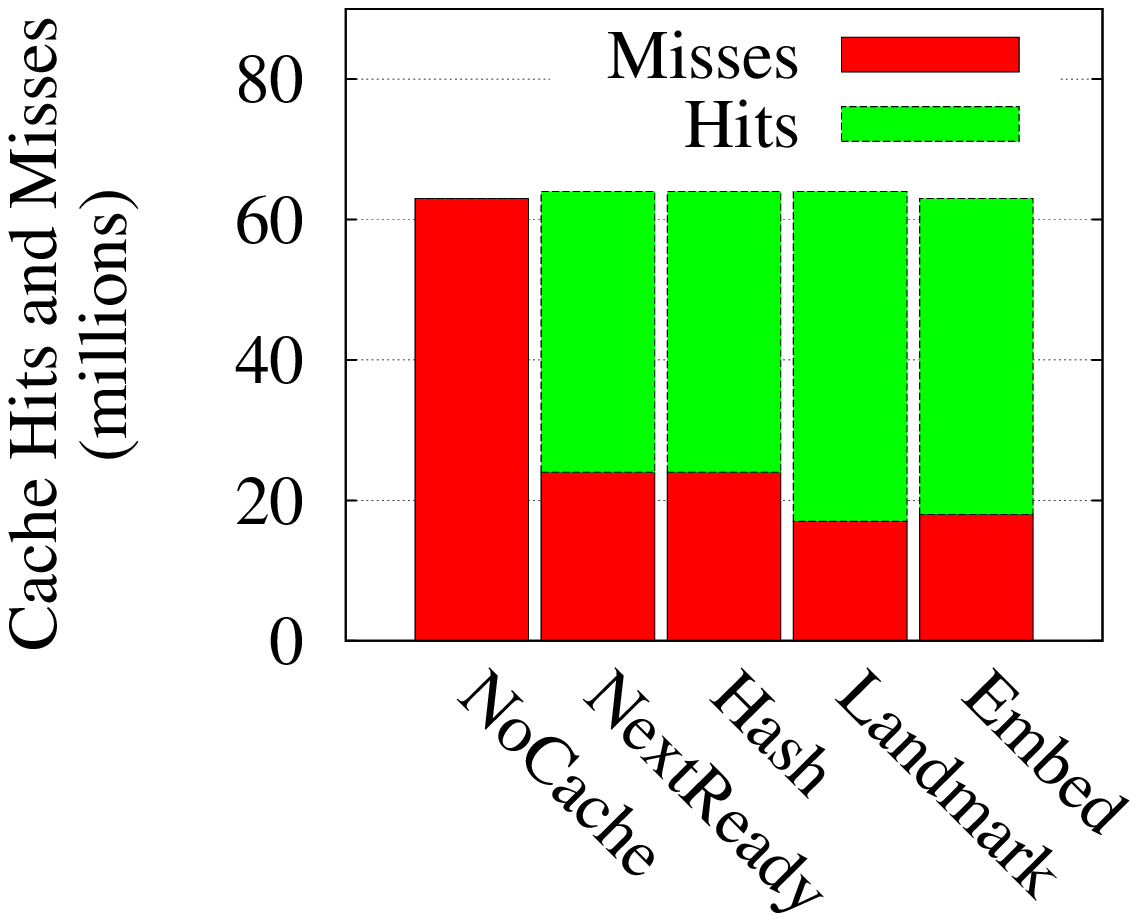}
\label{fig:hitmiss_1hop_web}
}
\subfigure [\small Cache hits and misses \newline  \hspace*{1.5em} (2-hop hotspot)] {
\includegraphics[scale=0.34]{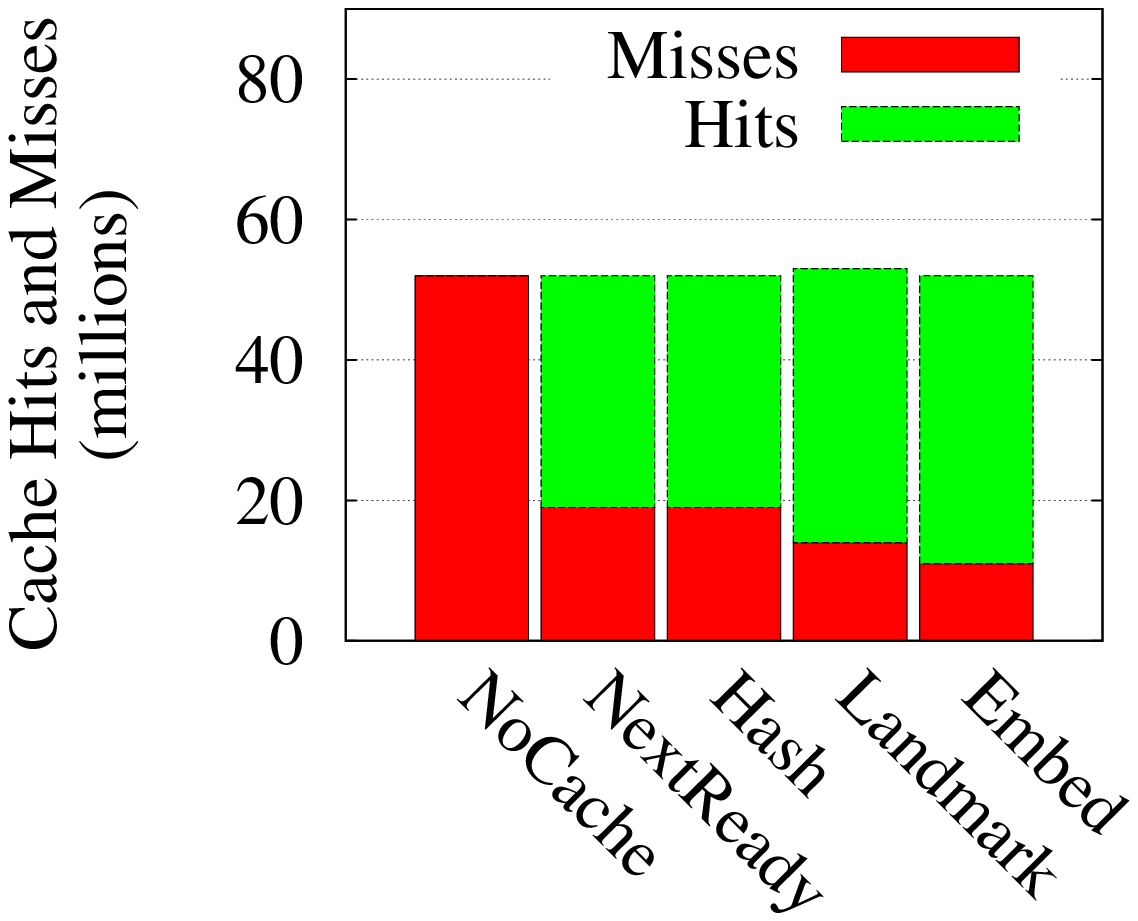}
\label{fig:hitmiss_2hop_web}
}
\vspace{-3mm}
\caption{\small Efficiency: $r$-hop hotspot, 2-hop traversal workload. We vary r = 1, 2. {\em WebGraph}.}
\label{fig:web_3hop}
\vspace{-5mm}
\end{figure*}
\begin{figure*}[tb!]
\centering
\subfigure [\small 1-hop traversal] {
\includegraphics[scale=0.34]{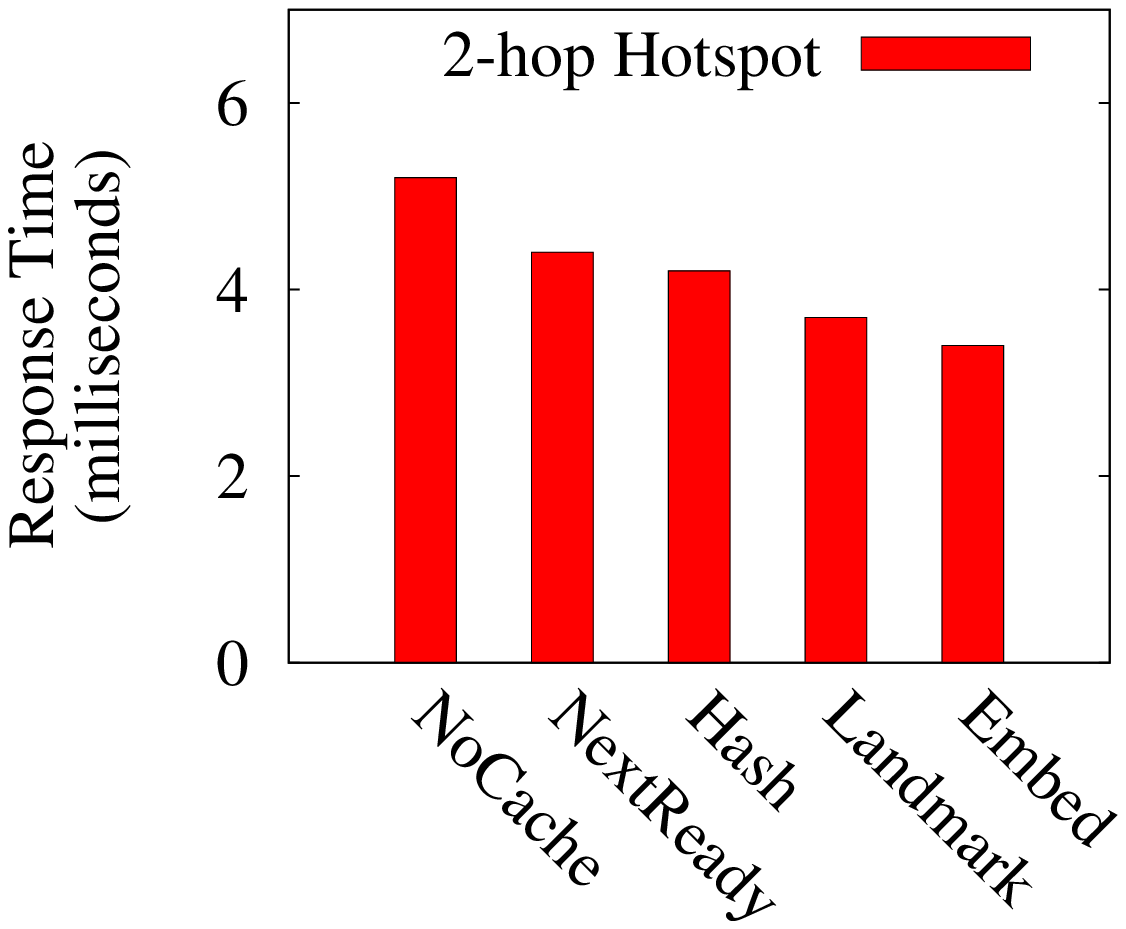}
\label{fig:response_time_web2}
}
\subfigure [\small 2-hop traversal] {
\includegraphics[scale=0.34]{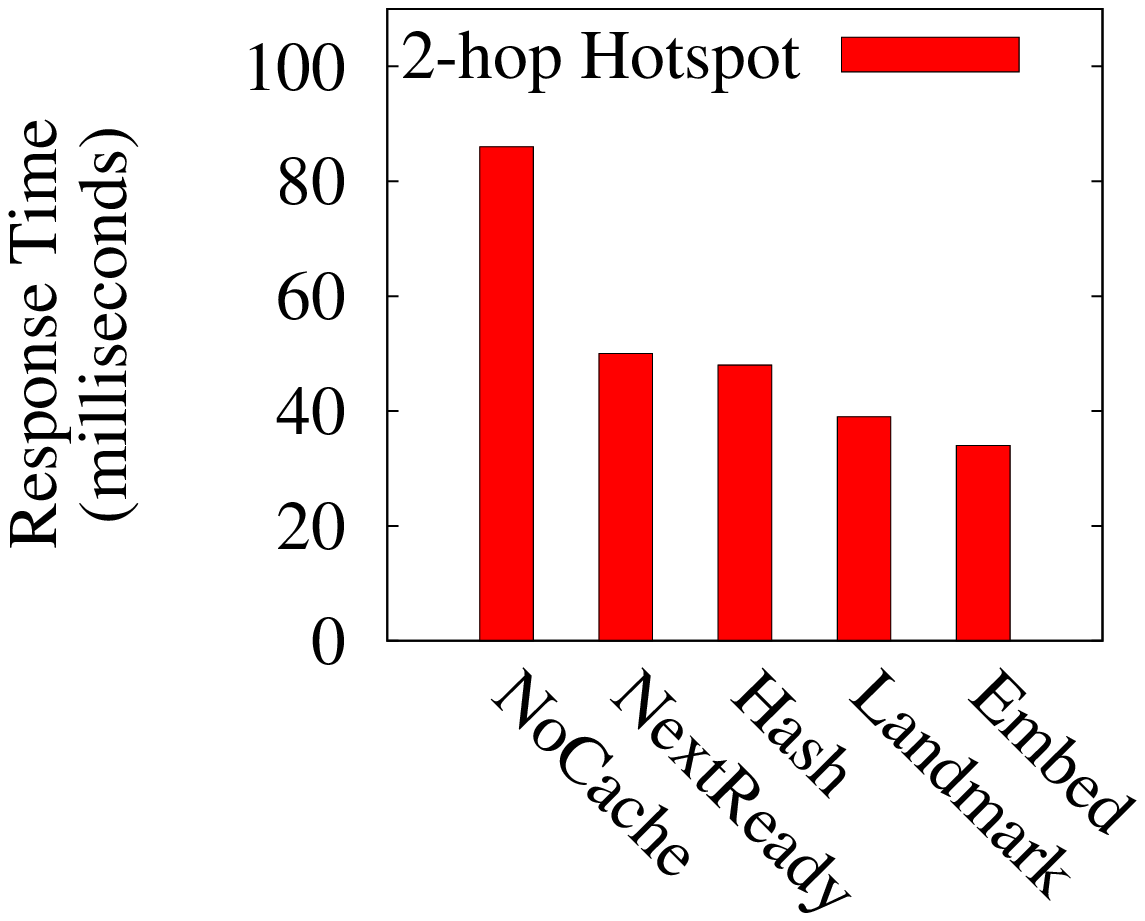}
\label{fig:response_time_web2}
}
\subfigure [\small 3-hop traversal] {
\includegraphics[scale=0.34]{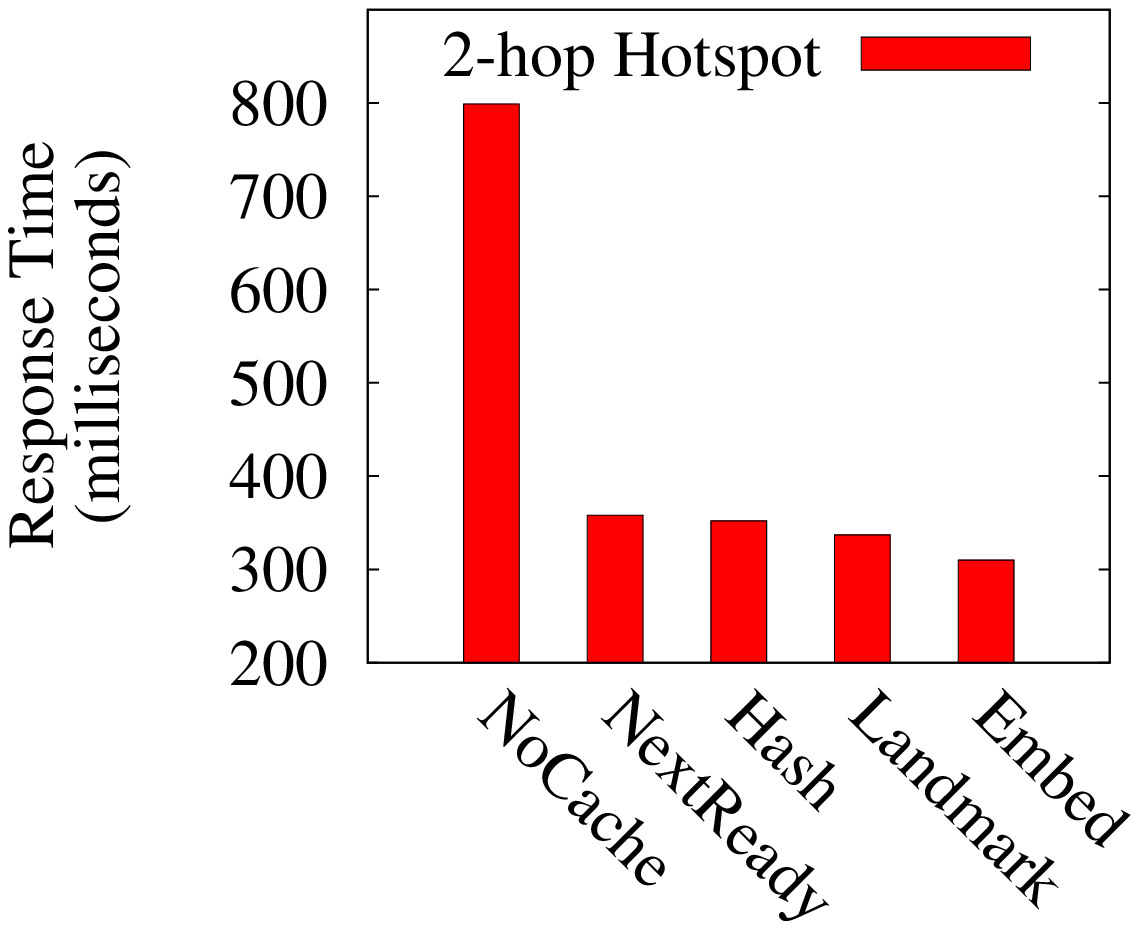}
\label{fig:response_time_web4}
}
\vspace{-3mm}
\caption{\small Efficiency: $2$-hop hotspot, h-hop traversal workload. We vary h = 1$\sim$3. {\em WebGraph}.}
\label{fig:web_h_hop}
\vspace{-5mm}
\end{figure*}

\vspace{-2mm}
\subsection{Sensitivity Analysis}
\label{sec:sensitivity}
We analyze the impact of various parameters, e.g., number of landmarks, their minimum separation, embedding
dimensionality, load factor, and smoothing factor ($\alpha$). We find that {\sf gRouting}
is more sensitive towards load factor (due to {\em query stealing}) and embedding dimensionality, compared to other parameters.
In all figures, we also present the response time (or, throughput)
of our best baseline, which is hash routing, for an effective comparison.

\spara{Load Factor:} This parameter impacts both of our smart routing schemes. We find from Equations 3 and 7
that smaller values of load factor diminish the impact of ``smart'' routing (i.e., landmarks and node co-ordinates),
instead queries will be routed to the processor having the minimum workload. On the other hand, higher values of load factor reduces the impact
of load balancing (i.e., query stealing) --- queries would be routed solely based on landmarks and node co-ordinates.
Therefore, in these experiments, we expect that the throughput will initially increase with higher values of load factor,
until it reaches a maximum, and then it would start decreasing. Indeed, it can be observed in Figure~\ref{fig:load} that
with load factor between 10$\sim$20, the best throughput is achieved.

\spara{Smoothing Parameter:} The smoothing parameter ($\alpha$) effects only embed routing, and
it can take values from 0 to 1, indicating how much new queries affect the query processor's expected
moving average ({\sf EMA}). A value of 1 would mean that the processor's {\sf EMA} would be equal to the
last query's co-ordinates, while a value of 0 would cause the processor's {\sf EMA} to retain
its initial value. We find in Figure~\ref{fig:alpha} that the response time for embed
routing reduces when $\alpha \in \{0.25, 0.75\}$.

\spara{Embedding Dimensionality:} We consider the performance implications of the number of dimensions on embed routing.
For these experiments, we create several embeddings, with dimensionality from 2 to 30. While the relative error in distance between node pairs decreases
with higher dimensions, it almost saturates after 10 dimensions (Figure~\ref{fig:dim_error}). On the other hand, we observe that the average response
time reduces until dimension 10, and then it slowly increases with more dimensions (Figure~\ref{fig:dim_response}). This is
because with higher dimensions, we reduce the distance prediction error, thereby correctly routing the queries and getting more cache hits. However,
a large number of dimensions also increases the routing decision making time at the router. Hence, the least response time is
achieved at dimensionality 10.

\spara{Number of Landmarks:} Figure~\ref{fig:no_landmark} shows the variation of query response time with the number of landmark nodes.
Both our smart routing strategies get affected by the number of landmarks --- generally
the more, the better. For example, we find that with 96 landmarks, there is a significant reduction in response time for embed routing.
We also recall that in the preprocessing phase of smart routings, one needs to compute the distance of every landmark to all nodes,
and thus, preprocessing time increases with more landmarks. Hence, we perform a trade-off between query response time and preprocessing time, thereby
setting the optimal number of landmarks as 96 in our experiments.

We also find that the distance between landmarks does not have a significant influence on response time, with embed's best performance at 3-hops separation
and landmark's at 4-hops separation (Figure~\ref{fig:separation_landmark}).
\vspace{-2mm}
\subsection{Varying Query Parameters ($r$, $h$)}
\label{sec:exp_q_parameter}
We analyze the impact of $r$ and $h$ in $r$-hop hotspot, $h$-hop traversal workloads.
In Figure~\ref{fig:web_3hop}, we vary $r$ as $1$ and $2$, while setting $h$ as $2$.
The average response times over {\em Webgraph} are shown in Figure~\ref{fig:response_time_web},
and the corresponding numbers of cache hits and misses are illustrated in Figures~\ref{fig:hitmiss_1hop_web}
and \ref{fig:hitmiss_2hop_web}. For both 1-hop and 2-hop hotspot queries,
we observe that our smart routing techniques, landmark and embed, outperform the
baseline routing algorithms. This is because smart routing schemes
are designed to capture topology-aware locality; and hence, make
better use of the processors' cache contents. In Figures 14(b) and
14(c), one may note that the smart routing schemes are indeed obtaining
more cache hits as compared to that of baseline routings.

Next, we run our experiments by setting $r$=2 and varying $h$ as $1$, $2$ and $3$.
As depicted in Figure~\ref{fig:web_h_hop}, the results for 1-hop traversal and 3-hop traversal
are exactly similar to what we found earlier for 2-hop traversal.
In addition, for 3-hop traversal (Figure~\ref{fig:response_time_web4}), the difference between smart routing
schemes and baselines has diminished compared to that for 1-hop and 2-hop traversal queries. Smart routing schemes
still obtain more cache hits per query and have lower response times, e.g., 15\% lower than baselines. However,
due to 3-hop traversals, these queries process a larger amount of data. For example,
the average 3-hop neighborhood size is 367K nodes.
This means that computation occupies a significant part of the overall response time, reducing the relative impact of
smart routing obtained via cache hits.
\vspace{-2mm}
\subsection{Results on Other Datasets}
\label{sec:exp_datasets}
\begin{figure}[tb!]
\centering
\subfigure [\small {\em Memetracker}]{
\includegraphics[scale=0.29]{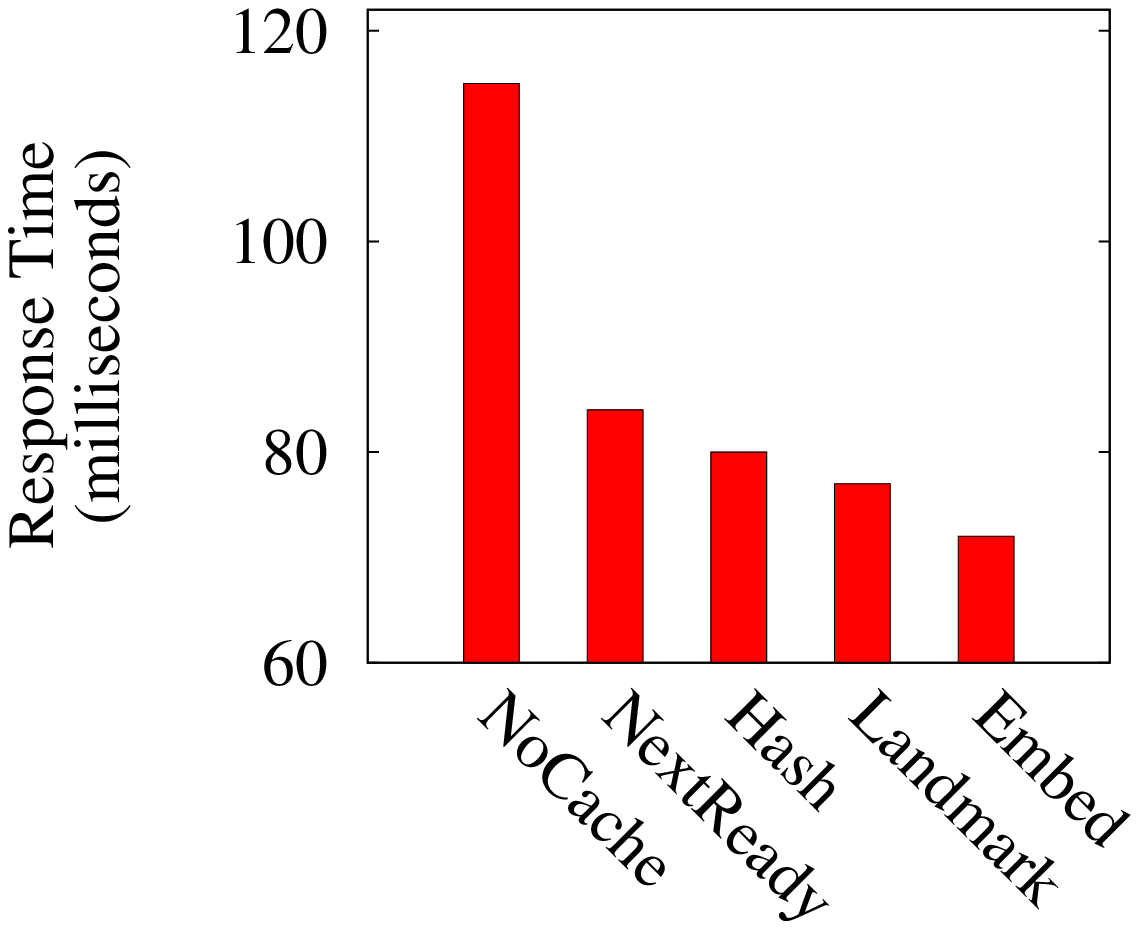}
\label{fig:response_meme}
}
\subfigure [\small {\em Friendster}] {
\includegraphics[scale=0.29]{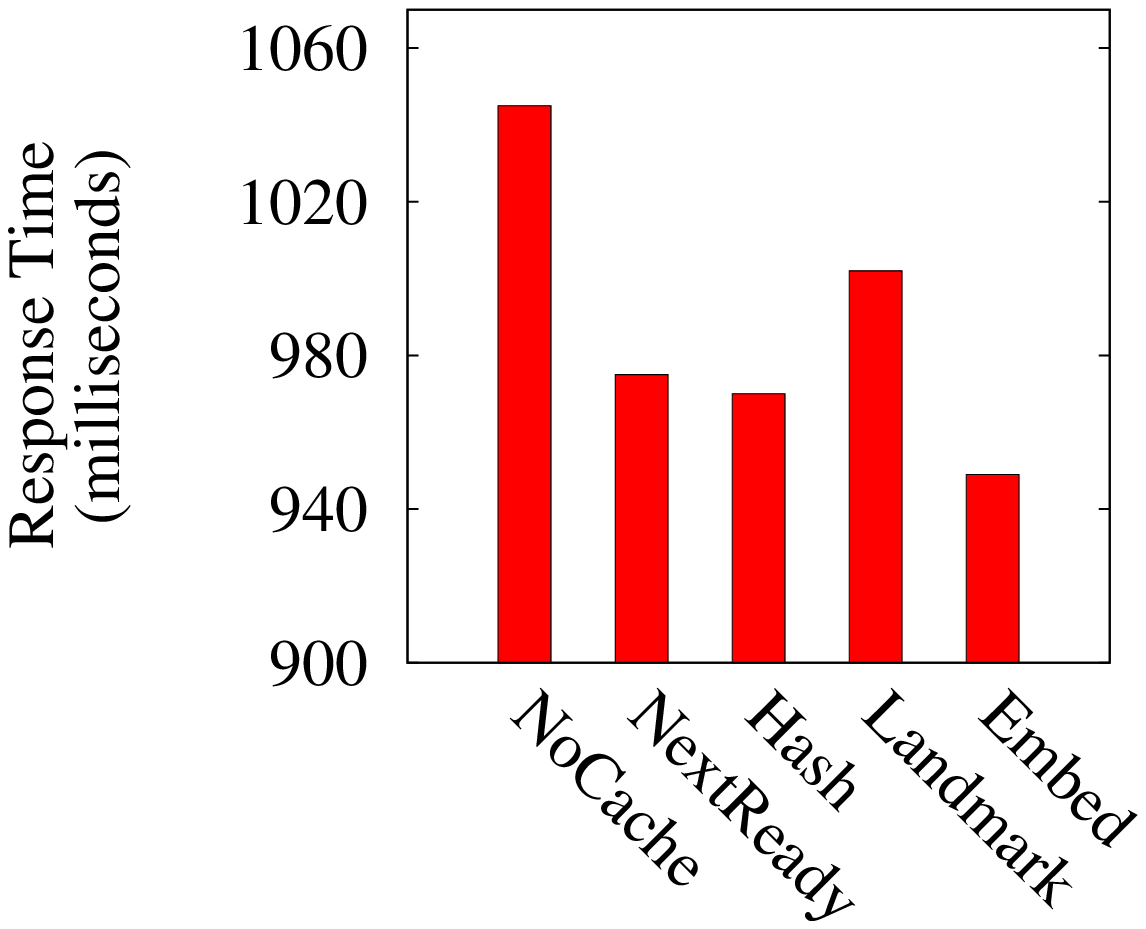}
\label{fig:response_friendster}
}
\vspace{-5mm}
\caption{\small Efficiency: {\em Memetracker}, {\em Friendster}, $2$-hop hotspot, 2-hop traversal workload}
\label{fig:efficiency_other}
\vspace{-6mm}
\end{figure}

We investigate how {\sf gRouting} performs over other graphs, e.g., {\em Memetracker} and {\em Friendster}.
We present the average response time with 2-hop hotspot, 2-hop traversal queries in Figure~\ref{fig:efficiency_other}.
The results with {\em Memetracker} are similar to those of our earlier results over {\em Webgraph}. In particular, baseline routing techniques
reduce the response time by 30\% compared to no-cache scheme, and smart routing algorithms further reduce it by 10\% in comparison with baseline routings (Figure~\ref{fig:response_meme}).

In case of {\em Friendster} (Figure~\ref{fig:response_friendster}), baseline routings reduce the response time by 7\% compared to no-cache scheme,
and smart routing techniques reduce it by another 3\% in comparison with baseline routings. Our investigation further reveals that the average 2-hop neighborhood size is 0.3M for this dataset, compared to 52K in case of {\em Webgraph}. Therefore, computation at the processing tier occupies a significant fraction of the overall response time. Besides, we find that the overlap across 2-hop neighborhoods for queries from the same hotspot region is lower than that of {\em Webgraph}. Thus caching becomes less effective here. Nevertheless, our embed smart routing reduces the response time by 40 ms (about 3\%) as compared to baseline routing schemes.

%% file: related.tex
\vspace{-1mm}
\section{related work}
\label{sec:related}
\vspace{-1mm}
We studied {\em smart query routing} for distributed graph querying --- a problem for which we are not aware of any prior
work. In the following we, however, provide
a brief overview of work in neighborhood areas.

\vspace{-1mm}
\spara{Landmarks and Graph Embedding.} Kleinberg et. al. \cite{KSW09} discussed the problem of approximating graph
distances using a small set of beacons (i.e., landmarks). Landmarks were also used in path finding, shortest path
estimation, as well as in estimating network properties \cite{RMJ06,AIY13,QCCY12}.
Graph embedding algorithms  \cite{ZSWZZ10}, on the other hand, place nodes at points on some surface such that
the inherent graph structure is preserved. Traditionally, spectral methods such as principal component analysis (PCA)
have been applied to many graph embedding and dimensionality reduction tasks, where the underlying manifold is a linear
subspace. For nonlinear cases, many recent techniques, e.g., multidimensional scaling (MDS), locally linear embedding (LLE),
Laplacian eigenmaps (LEM), Isomap, semidefinite embedding (SDE), minimum volume embedding (MVE), and structure preserving
embedding (SPE) are proposed. In the past, graph embedding schemes were employed in internet routing,
such as predicting internet network distances and estimating minimum round trip time between hosts \cite{DCKM04}.
To the best of our knowledge, ours is the first study that applies graph embedding and landmark techniques to
design effective routing algorithms for distributed graph querying.

\vspace{-2mm}
\spara{Graph Partitioning, Re-partitioning, Replication.} The balanced, minimum-edge-cut graph partitioning
divides a graph into $k$ partitions such that each partition contains same number of nodes,
and the number of cut-edges is minimized. Even for $k=2$, the problem is \NP-hard, and there is no approximation algorithm with a constant approximation ratio unless \onlyP=\NP \cite{NKDC15}.
Therefore, efforts were made in developing polynomial-time heuristics --- {\sf METIS}, {\sf Chaco}, {\sf SCOTCH}, to name a few.
More sophisticated graph partitioning schemes were also proposed, e.g., node-cut \cite{GLGBG12}, complementary partitioning \cite{YYZK12},
and label propagation \cite{WXSW14}, among many others.

Graph re-partitioning is critical for online queries, since the graph topology and workload change over time
\cite{MD12}. The methods in \cite{YYZK12,NKDC15,KAAJWK13} perform re-partitioning
based on past workloads. Incremental techniques were proposed for partitioning dynamic and stream graphs \cite{ZLC16,VCLM13,SG12}.
However, online monitoring of the load imbalance and graph updates, dynamically modifying the graph partitions,
and migrating data across servers are expensive. With the proposed embed routing, we bypass
these expensive graph partitioning and re-partitioning challenges to the existing cache replacement policy of the query processors.

Replication was used for graph partitioning, re-partitioning, load balancing, and
fault tolerance. In earlier works, \cite{PESYLCR10,HAR11} proposed one extreme version by replicating the graph
sufficiently so that, for every node in the graph, all of its neighbors are present locally.
Mondal et. al.  designed an overlapping graph re-partitioning scheme \cite{MD12}, which updates its
partitions based on the past read/ write patterns. Huang et. al. \cite{HA16} designed a lightweight
re- partitioning and replication scheme considering access locality, fault tolerance, and dynamic
updates. While we also replicate the graph data at query processors' cache in an overlapping manner,
we {\em only replicate the active regions} of the graph based on most recent workloads.
More importantly, unlike \cite{MD12,HA16} we do not explicitly run any graph replication strategy at our processors or storage servers.
Instead, our {\em smart routing algorithms} automatically perform replications at processors' cache.

\vspace{-2mm}
\spara{Graph Caching, De-coupling, Multi-Query Optimization.} {\sf Facebook} uses a fast caching layer, {\sf Memcached} on top of a graph database to scale the
performance of graph querying \cite{NFGKLLMPPSSV13}. Graph-structure-aware and workload-adaptive caching techniques were
also proposed, e.g., \cite{ACCKU15, PTKK15}. There are other works on view-based graph query answering \cite{FWW14} and multi-query optimizations \cite{LKDL12}.
Unlike ours, these approaches require the workload to be known in advance.

Recently, Shalita et. al. \cite{SKKSPAKS16} employed decoupling for an optimal assignment of HTTP requests over a distributed graph storage.
First, they perform a static partition of the graph in storage servers based on its structure and co-access patterns. Next, they
find past workloads on each partition, and dynamically assign these partitions to query processors such that load balancing can be achieved.
While their decoupling principle and dynamic assignment at query processors are similar to ours, they still explicitly perform a sophisticated
graph partitioning at the storage servers, and update such partitions in an offline manner. In contrast, our smart routing algorithms
automatically partition the active regions of the graph in a dynamic manner and store them in the query processors' cache, thereby
achieving both load balancing and improved cache hit rates.

\vspace{-1mm}
\spara{Large-Scale Graph Processing Systems.} Distributed graph processing systems can be classified based on application support:
{\bf (1)} systems for offline analytics and {\bf (2)} systems for online querying. Offline analytic systems
perform iterative, batch processing over the graph by following a node/edge-centric scatter gather model.
Examples of such computations include {\sf PageRank}, diameter of a graph, and strongly connected components.
{\sf Pregel} \cite{MABDHLC10} and {\sf PowerGraph} \cite{GLGBG12} are representatives of offline graph processing systems.

Offline graph processing systems are often not suitable for online graph queries \cite{SWL13}.
This is because online graph querying requires fast response time, and these queries explore parts of the graph.
Examples of online graph queries are $h$-hop neighborhood search, shortest-path finding, sub-graph matching, and {\sf SPARQL} queries.
As discussed in \cite{SWL13}, online graph querying systems must facilitate graph traversal over a region of the graph for faster query answering.
{\sf Horton} \cite{SEHM13} and {\sf Ligra} \cite{SB13} are examples of online graph querying systems.
{\sf Trinity} \cite{SWL13} supports both offline graph analytics and online graph queries.
The de-coupled architecture and our designed smart routing algorithms,
being generic, can benefit the existing online graph querying systems.

%% file: conclusion.tex
\vspace{-2mm}
\section{CONCLUSIONS}
\label{sec:conclusion}
\vspace{-1mm}
We studied $h$-hop traversal queries -- a generalized
form of various online graph queries that access a small region of the
graph, and require fast response time. To answer such queries with low latency
and high throughput, we follow the principle of decoupling query processors from graph storage.
Our work emphasized {\em less} on the requirements for an expensive graph partitioning and re-partitioning technique,
instead we developed smart query routing strategies for effectively leveraging the query processors'
cache contents, thereby improving the throughput and
reducing latency of distributed graph querying.
In addition to workload balancing and deployment flexibility, {\sf gRouting} is able to
provide linear scalability in throughput with more number of query processors,
works well in the presence of query hotspots, and is also adaptive to workload changes and graph updates.

In future, enabling modern query processing techniques and hardware trends,
e.g., parallel, collaborative scans in main memory could be interesting to support a larger set
of queries. It would also be worth investigating transaction workloads such as
streaming graph updates and consistency criteria in a decoupled system.

\vspace{-2mm}
\section{Acknowledgement} Arijit Khan is supported by MOE Tier-1 M401020000 and NTU M4081678. Any opinions, findings, and conclusions in this publication are those of the authors and do not necessarily reflect the views of the funding agencies.

%% file: gRouting.bbl

\begin{thebibliography}{00}


\ifx \showCODEN    \undefined \def \showCODEN     #1{\unskip}     \fi
\ifx \showDOI      \undefined \def \showDOI       #1{{\tt DOI:}\penalty0{#1}\ }
  \fi
\ifx \showISBNx    \undefined \def \showISBNx     #1{\unskip}     \fi
\ifx \showISBNxiii \undefined \def \showISBNxiii  #1{\unskip}     \fi
\ifx \showISSN     \undefined \def \showISSN      #1{\unskip}     \fi
\ifx \showLCCN     \undefined \def \showLCCN      #1{\unskip}     \fi
\ifx \shownote     \undefined \def \shownote      #1{#1}          \fi
\ifx \showarticletitle \undefined \def \showarticletitle #1{#1}   \fi
\ifx \showURL      \undefined \def \showURL       {\relax}        \fi
\providecommand\bibfield[2]{#2}
\providecommand\bibinfo[2]{#2}
\providecommand\natexlab[1]{#1}
\providecommand\showeprint[2][]{arXiv:#2}

\bibitem[\protect\citeauthoryear{Akiba, Iwata, and Yoshida}{Akiba
  et~al\mbox{.}}{2013}]%
        {AIY13}
\bibfield{author}{\bibinfo{person}{T. Akiba}, \bibinfo{person}{Y. Iwata}, {and}
  \bibinfo{person}{Y. Yoshida}.} \bibinfo{year}{2013}\natexlab{}.
\newblock \showarticletitle{{Fast Exact Shortest-path Distance Queries on Large
  Networks by Pruned Landmark Labeling}}. In \bibinfo{booktitle}{{\em SIGMOD}}.
\newblock


\bibitem[\protect\citeauthoryear{Aksu, Canim, Chang, Korpeoglu, and
  Ulusoy}{Aksu et~al\mbox{.}}{2015}]%
        {ACCKU15}
\bibfield{author}{\bibinfo{person}{H. Aksu}, \bibinfo{person}{M. Canim},
  \bibinfo{person}{Y.{-}C. Chang}, \bibinfo{person}{I. Korpeoglu}, {and}
  \bibinfo{person}{{\"{O}}. Ulusoy}.} \bibinfo{year}{2015}\natexlab{}.
\newblock \showarticletitle{{Graph Aware Caching Policy for Distributed Graph
  Stores}}. In \bibinfo{booktitle}{{\em IC2E}}.
\newblock


\bibitem[\protect\citeauthoryear{Binnig, Crotty, Galakatos, Kraska, and
  Zamanian}{Binnig et~al\mbox{.}}{2016}]%
        {BCGKZ16}
\bibfield{author}{\bibinfo{person}{C. Binnig}, \bibinfo{person}{A. Crotty},
  \bibinfo{person}{A. Galakatos}, \bibinfo{person}{T. Kraska}, {and}
  \bibinfo{person}{E. Zamanian}.} \bibinfo{year}{2016}\natexlab{}.
\newblock \showarticletitle{{The End of Slow Networks: It's Time for a
  Redesign}}.
\newblock \bibinfo{journal}{{\em PVLDB\/}} \bibinfo{volume}{9},
  \bibinfo{number}{7} (\bibinfo{year}{2016}), \bibinfo{pages}{528--539}.
\newblock


\bibitem[\protect\citeauthoryear{Dabek, Cox, Kaashoek, and Morris}{Dabek
  et~al\mbox{.}}{2004}]%
        {DCKM04}
\bibfield{author}{\bibinfo{person}{F. Dabek}, \bibinfo{person}{R. Cox},
  \bibinfo{person}{F. Kaashoek}, {and} \bibinfo{person}{R. Morris}.}
  \bibinfo{year}{2004}\natexlab{}.
\newblock \showarticletitle{{Vivaldi: A Decentralized Network Coordinate
  System}}. In \bibinfo{booktitle}{{\em SIGCOMM}}.
\newblock


\bibitem[\protect\citeauthoryear{Fan, Wang, and Wu}{Fan et~al\mbox{.}}{2014}]%
        {FWW14}
\bibfield{author}{\bibinfo{person}{W. Fan}, \bibinfo{person}{X. Wang}, {and}
  \bibinfo{person}{Y. Wu}.} \bibinfo{year}{2014}\natexlab{}.
\newblock \showarticletitle{{Answering Graph Pattern Queries using Views}}. In
  \bibinfo{booktitle}{{\em ICDE}}.
\newblock


\bibitem[\protect\citeauthoryear{Gonzalez, Low, Gu, Bickson, and
  Guestrin}{Gonzalez et~al\mbox{.}}{2012}]%
        {GLGBG12}
\bibfield{author}{\bibinfo{person}{J.~E. Gonzalez}, \bibinfo{person}{Y. Low},
  \bibinfo{person}{H. Gu}, \bibinfo{person}{D. Bickson}, {and}
  \bibinfo{person}{C. Guestrin}.} \bibinfo{year}{2012}\natexlab{}.
\newblock \showarticletitle{{PowerGraph: Distributed Graph-parallel Computation
  on Natural Graphs}}. In \bibinfo{booktitle}{{\em OSDI}}.
\newblock


\bibitem[\protect\citeauthoryear{Huang and Abadi}{Huang and Abadi}{2016}]%
        {HA16}
\bibfield{author}{\bibinfo{person}{J. Huang} {and} \bibinfo{person}{D.~J.
  Abadi}.} \bibinfo{year}{2016}\natexlab{}.
\newblock \showarticletitle{{Leopard: Lightweight Edge-oriented Partitioning
  and Replication for Dynamic Graphs}}.
\newblock \bibinfo{journal}{{\em PVLDB\/}} \bibinfo{volume}{9},
  \bibinfo{number}{7} (\bibinfo{year}{2016}), \bibinfo{pages}{540--551}.
\newblock


\bibitem[\protect\citeauthoryear{Huang, Abadi, and Ren}{Huang
  et~al\mbox{.}}{2011}]%
        {HAR11}
\bibfield{author}{\bibinfo{person}{J. Huang}, \bibinfo{person}{D.~J. Abadi},
  {and} \bibinfo{person}{K. Ren}.} \bibinfo{year}{2011}\natexlab{}.
\newblock \showarticletitle{{Scalable SPARQL Querying of Large RDF Graphs}}.
\newblock \bibinfo{journal}{{\em PVLDB\/}} \bibinfo{volume}{4},
  \bibinfo{number}{11} (\bibinfo{year}{2011}), \bibinfo{pages}{1123--1134}.
\newblock


\bibitem[\protect\citeauthoryear{Karypis}{Karypis}{2011}]%
        {KK98}
\bibfield{author}{\bibinfo{person}{G. Karypis}.}
  \bibinfo{year}{2011}\natexlab{}.
\newblock \showarticletitle{{METIS and ParMETIS}}.
\newblock In \bibinfo{booktitle}{{\em Encyclopedia of Parallel Computing}}.
  \bibinfo{publisher}{Springer}.
\newblock


\bibitem[\protect\citeauthoryear{Khayyat, Awara, Alonazi, Jamjoom, Williams,
  and Kalnis}{Khayyat et~al\mbox{.}}{2013}]%
        {KAAJWK13}
\bibfield{author}{\bibinfo{person}{Z. Khayyat}, \bibinfo{person}{K. Awara},
  \bibinfo{person}{A. Alonazi}, \bibinfo{person}{H. Jamjoom},
  \bibinfo{person}{D. Williams}, {and} \bibinfo{person}{P. Kalnis}.}
  \bibinfo{year}{2013}\natexlab{}.
\newblock \showarticletitle{{Mizan: A System for Dynamic Load Balancing in
  Large-scale Graph Processing}}. In \bibinfo{booktitle}{{\em EuroSys}}.
\newblock


\bibitem[\protect\citeauthoryear{Kleinberg, Slivkins, and Wexler}{Kleinberg
  et~al\mbox{.}}{2009}]%
        {KSW09}
\bibfield{author}{\bibinfo{person}{J. Kleinberg}, \bibinfo{person}{A.
  Slivkins}, {and} \bibinfo{person}{T. Wexler}.}
  \bibinfo{year}{2009}\natexlab{}.
\newblock \showarticletitle{{Triangulation and Embedding Using Small Sets of
  Beacons}}.
\newblock \bibinfo{journal}{{\em J. ACM\/}} \bibinfo{volume}{56},
  \bibinfo{number}{6}, Article \bibinfo{articleno}{32} (\bibinfo{year}{2009}),
  \bibinfo{numpages}{32:1--32:37}~pages.
\newblock


\bibitem[\protect\citeauthoryear{Le, Kementsietsidis, Duan, and Li}{Le
  et~al\mbox{.}}{2012}]%
        {LKDL12}
\bibfield{author}{\bibinfo{person}{W. Le}, \bibinfo{person}{A.
  Kementsietsidis}, \bibinfo{person}{S. Duan}, {and} \bibinfo{person}{F. Li}.}
  \bibinfo{year}{2012}\natexlab{}.
\newblock \showarticletitle{{Scalable Multi-query Optimization for SPARQL}}. In
  \bibinfo{booktitle}{{\em ICDE}}.
\newblock


\bibitem[\protect\citeauthoryear{Loesing, Pilman, Etter, and Kossmann}{Loesing
  et~al\mbox{.}}{2015}]%
        {LPEK15}
\bibfield{author}{\bibinfo{person}{S. Loesing}, \bibinfo{person}{M. Pilman},
  \bibinfo{person}{T. Etter}, {and} \bibinfo{person}{D. Kossmann}.}
  \bibinfo{year}{2015}\natexlab{}.
\newblock \showarticletitle{{On the Design and Scalability of Distributed
  Shared-Data Databases}}. In \bibinfo{booktitle}{{\em SIGMOD}}.
\newblock


\bibitem[\protect\citeauthoryear{Malewicz, Austern, Bik, Dehnert, Horn, Leiser,
  and Czajkowski}{Malewicz et~al\mbox{.}}{2010}]%
        {MABDHLC10}
\bibfield{author}{\bibinfo{person}{G. Malewicz}, \bibinfo{person}{M.~H.
  Austern}, \bibinfo{person}{A.~J.~C. Bik}, \bibinfo{person}{J.~C. Dehnert},
  \bibinfo{person}{I. Horn}, \bibinfo{person}{N. Leiser}, {and}
  \bibinfo{person}{G. Czajkowski}.} \bibinfo{year}{2010}\natexlab{}.
\newblock \showarticletitle{{Pregel: A System for Large-scale Graph
  Processing}}. In \bibinfo{booktitle}{{\em SIGMOD}}.
\newblock


\bibitem[\protect\citeauthoryear{Mondal and Deshpande}{Mondal and
  Deshpande}{2012}]%
        {MD12}
\bibfield{author}{\bibinfo{person}{J. Mondal} {and} \bibinfo{person}{A.
  Deshpande}.} \bibinfo{year}{2012}\natexlab{}.
\newblock \showarticletitle{{Managing Large Dynamic Graphs Efficiently}}. In
  \bibinfo{booktitle}{{\em SIGMOD}}.
\newblock


\bibitem[\protect\citeauthoryear{Mondal and Deshpande}{Mondal and
  Deshpande}{2014}]%
        {MD14}
\bibfield{author}{\bibinfo{person}{J. Mondal} {and} \bibinfo{person}{A.
  Deshpande}.} \bibinfo{year}{2014}\natexlab{}.
\newblock \showarticletitle{{EAGr: Supporting Continuous Ego-centric Aggregate
  Queries over Large Dynamic Graphs}}. In \bibinfo{booktitle}{{\em SIGMOD}}.
\newblock


\bibitem[\protect\citeauthoryear{Nicoara, Kamali, Daudjee, and Chen}{Nicoara
  et~al\mbox{.}}{2015}]%
        {NKDC15}
\bibfield{author}{\bibinfo{person}{D. Nicoara}, \bibinfo{person}{S. Kamali},
  \bibinfo{person}{K. Daudjee}, {and} \bibinfo{person}{L. Chen}.}
  \bibinfo{year}{2015}\natexlab{}.
\newblock \showarticletitle{{Hermes: Dynamic Partitioning for Distributed
  Social Network Graph Databases}}. In \bibinfo{booktitle}{{\em EDBT}}.
\newblock


\bibitem[\protect\citeauthoryear{Nishtala, Fugal, Grimm, Kwiatkowski, Lee, Li,
  McElroy, Paleczny, Peek, Saab, Stafford, Tung, and Venkataramani}{Nishtala
  et~al\mbox{.}}{2013}]%
        {NFGKLLMPPSSV13}
\bibfield{author}{\bibinfo{person}{R. Nishtala}, \bibinfo{person}{H. Fugal},
  \bibinfo{person}{S. Grimm}, \bibinfo{person}{M. Kwiatkowski},
  \bibinfo{person}{H. Lee}, \bibinfo{person}{H.~C. Li}, \bibinfo{person}{R.
  McElroy}, \bibinfo{person}{M. Paleczny}, \bibinfo{person}{D. Peek},
  \bibinfo{person}{P. Saab}, \bibinfo{person}{D. Stafford}, \bibinfo{person}{T.
  Tung}, {and} \bibinfo{person}{V. Venkataramani}.}
  \bibinfo{year}{2013}\natexlab{}.
\newblock \showarticletitle{{Scaling Memcache at Facebook}}. In
  \bibinfo{booktitle}{{\em NSDI}}.
\newblock


\bibitem[\protect\citeauthoryear{Ousterhout, Agrawal, Erickson, Kozyrakis,
  Leverich, Mazi\`{e}res, Mitra, Narayanan, Parulkar, Rosenblum, Rumble,
  Stratmann, and Stutsman}{Ousterhout et~al\mbox{.}}{2010}]%
        {OAEK10}
\bibfield{author}{\bibinfo{person}{J. Ousterhout}, \bibinfo{person}{P.
  Agrawal}, \bibinfo{person}{D. Erickson}, \bibinfo{person}{C. Kozyrakis},
  \bibinfo{person}{J. Leverich}, \bibinfo{person}{D. Mazi\`{e}res},
  \bibinfo{person}{S. Mitra}, \bibinfo{person}{A. Narayanan},
  \bibinfo{person}{G. Parulkar}, \bibinfo{person}{M. Rosenblum},
  \bibinfo{person}{S.~M. Rumble}, \bibinfo{person}{E. Stratmann}, {and}
  \bibinfo{person}{R. Stutsman}.} \bibinfo{year}{2010}\natexlab{}.
\newblock \showarticletitle{{The Case for RAMClouds: Scalable High-performance
  Storage Entirely in DRAM}}.
\newblock \bibinfo{journal}{{\em SIGOPS Oper. Syst. Rev.\/}}
  \bibinfo{volume}{43}, \bibinfo{number}{4} (\bibinfo{year}{2010}),
  \bibinfo{pages}{92--105}.
\newblock


\bibitem[\protect\citeauthoryear{Papailiou, Tsoumakos, Karras, and
  Koziris}{Papailiou et~al\mbox{.}}{2015}]%
        {PTKK15}
\bibfield{author}{\bibinfo{person}{N. Papailiou}, \bibinfo{person}{D.
  Tsoumakos}, \bibinfo{person}{P. Karras}, {and} \bibinfo{person}{N. Koziris}.}
  \bibinfo{year}{2015}\natexlab{}.
\newblock \showarticletitle{{Graph-Aware, Workload-Adaptive SPARQL Query
  Caching}}. In \bibinfo{booktitle}{{\em SIGMOD}}.
\newblock


\bibitem[\protect\citeauthoryear{Pujol, Erramilli, Siganos, Yang, Laoutaris,
  Chhabra, and Rodriguez}{Pujol et~al\mbox{.}}{2010}]%
        {PESYLCR10}
\bibfield{author}{\bibinfo{person}{J.~M. Pujol}, \bibinfo{person}{V.
  Erramilli}, \bibinfo{person}{G. Siganos}, \bibinfo{person}{X. Yang},
  \bibinfo{person}{N. Laoutaris}, \bibinfo{person}{P. Chhabra}, {and}
  \bibinfo{person}{P. Rodriguez}.} \bibinfo{year}{2010}\natexlab{}.
\newblock \showarticletitle{{The Little Engine(s) That Could: Scaling Online
  Social Networks}}. In \bibinfo{booktitle}{{\em SIGCOMM}}.
\newblock


\bibitem[\protect\citeauthoryear{Qiao, Cheng, Chang, and Yu}{Qiao
  et~al\mbox{.}}{2012}]%
        {QCCY12}
\bibfield{author}{\bibinfo{person}{M. Qiao}, \bibinfo{person}{H. Cheng},
  \bibinfo{person}{L. Chang}, {and} \bibinfo{person}{J.~X. Yu}.}
  \bibinfo{year}{2012}\natexlab{}.
\newblock \showarticletitle{{Approximate Shortest Distance Computing: A
  Query-Dependent Local Landmark Scheme}}. In \bibinfo{booktitle}{{\em ICDE}}.
\newblock


\bibitem[\protect\citeauthoryear{Rattigan, Maier, and Jensen}{Rattigan
  et~al\mbox{.}}{2006}]%
        {RMJ06}
\bibfield{author}{\bibinfo{person}{M.~J. Rattigan}, \bibinfo{person}{M.~E.
  Maier}, {and} \bibinfo{person}{D. Jensen}.} \bibinfo{year}{2006}\natexlab{}.
\newblock \showarticletitle{{Using Structure Indices for Efficient
  Approximation of Network Properties}}. In \bibinfo{booktitle}{{\em KDD}}.
\newblock


\bibitem[\protect\citeauthoryear{Roy, Bindschaelder, Malicevic, and
  Zwaenepoel}{Roy et~al\mbox{.}}{2015}]%
        {RBMZ15}
\bibfield{author}{\bibinfo{person}{A. Roy}, \bibinfo{person}{L. Bindschaelder},
  \bibinfo{person}{J. Malicevic}, {and} \bibinfo{person}{W. Zwaenepoel}.}
  \bibinfo{year}{2015}\natexlab{}.
\newblock \showarticletitle{{Chaos: Scale-out Graph Processing from Secondary
  Storage}}. In \bibinfo{booktitle}{{\em SOSP}}.
\newblock


\bibitem[\protect\citeauthoryear{Sarwat, Elnikety, He, and Mokbel}{Sarwat
  et~al\mbox{.}}{2013}]%
        {SEHM13}
\bibfield{author}{\bibinfo{person}{M. Sarwat}, \bibinfo{person}{S. Elnikety},
  \bibinfo{person}{Y. He}, {and} \bibinfo{person}{M.~F. Mokbel}.}
  \bibinfo{year}{2013}\natexlab{}.
\newblock \showarticletitle{{Horton+: A Distributed System for Processing
  Declarative Reachability Queries over Partitioned Graphs}}.
\newblock \bibinfo{journal}{{\em PVLDB\/}} \bibinfo{volume}{6},
  \bibinfo{number}{14} (\bibinfo{year}{2013}), \bibinfo{pages}{1918--1929}.
\newblock


\bibitem[\protect\citeauthoryear{Shalita, Karrer, Kabiljo, Sharma, Presta,
  Adcock, Kllapi, and Stumm}{Shalita et~al\mbox{.}}{2016}]%
        {SKKSPAKS16}
\bibfield{author}{\bibinfo{person}{A. Shalita}, \bibinfo{person}{B. Karrer},
  \bibinfo{person}{I. Kabiljo}, \bibinfo{person}{A. Sharma},
  \bibinfo{person}{A. Presta}, \bibinfo{person}{A. Adcock}, \bibinfo{person}{H.
  Kllapi}, {and} \bibinfo{person}{M. Stumm}.} \bibinfo{year}{2016}\natexlab{}.
\newblock \showarticletitle{{Social Hash: An Assignment Framework for
  Optimizing Distributed Systems Operations on Social Networks}}. In
  \bibinfo{booktitle}{{\em NSDI}}.
\newblock


\bibitem[\protect\citeauthoryear{Shao, Wang, and Li}{Shao
  et~al\mbox{.}}{2013}]%
        {SWL13}
\bibfield{author}{\bibinfo{person}{B. Shao}, \bibinfo{person}{H. Wang}, {and}
  \bibinfo{person}{Y. Li}.} \bibinfo{year}{2013}\natexlab{}.
\newblock \showarticletitle{{Trinity: A Distributed Graph Engine on a Memory
  Cloud}}. In \bibinfo{booktitle}{{\em SIGMOD}}.
\newblock


\bibitem[\protect\citeauthoryear{Shun and Blelloch}{Shun and Blelloch}{2013}]%
        {SB13}
\bibfield{author}{\bibinfo{person}{J. Shun} {and} \bibinfo{person}{G.~E.
  Blelloch}.} \bibinfo{year}{2013}\natexlab{}.
\newblock \showarticletitle{{Ligra: A Lightweight Graph Processing Framework
  for Shared Memory}}. In \bibinfo{booktitle}{{\em PPoPP}}.
\newblock


\bibitem[\protect\citeauthoryear{Shute, Vingralek, Samwel, Handy, Whipkey,
  Rollins, Oancea, Littlefield, Menestrina, Ellner, Cieslewicz, Rae, Stancescu,
  and Apte}{Shute et~al\mbox{.}}{2013}]%
        {SVSH13}
\bibfield{author}{\bibinfo{person}{J. Shute}, \bibinfo{person}{R. Vingralek},
  \bibinfo{person}{B. Samwel}, \bibinfo{person}{B. Handy}, \bibinfo{person}{C.
  Whipkey}, \bibinfo{person}{E. Rollins}, \bibinfo{person}{M. Oancea},
  \bibinfo{person}{K. Littlefield}, \bibinfo{person}{D. Menestrina},
  \bibinfo{person}{S. Ellner}, \bibinfo{person}{J. Cieslewicz},
  \bibinfo{person}{I. Rae}, \bibinfo{person}{T. Stancescu}, {and}
  \bibinfo{person}{H. Apte}.} \bibinfo{year}{2013}\natexlab{}.
\newblock \showarticletitle{{F1: A Distributed SQL Database That Scales}}.
\newblock \bibinfo{journal}{{\em PVLDB\/}} \bibinfo{volume}{6},
  \bibinfo{number}{11} (\bibinfo{year}{2013}), \bibinfo{pages}{1068--1079}.
\newblock


\bibitem[\protect\citeauthoryear{Stanton and Kliot}{Stanton and Kliot}{2012}]%
        {SG12}
\bibfield{author}{\bibinfo{person}{I. Stanton} {and} \bibinfo{person}{G.
  Kliot}.} \bibinfo{year}{2012}\natexlab{}.
\newblock \showarticletitle{{Streaming Graph Partitioning for Large Distributed
  Graphs}}. In \bibinfo{booktitle}{{\em KDD}}.
\newblock


\bibitem[\protect\citeauthoryear{Tretyakov, A.-Cervantes, G.-Banuelos, Vilo,
  and Dumas}{Tretyakov et~al\mbox{.}}{2011}]%
        {TAGVD11}
\bibfield{author}{\bibinfo{person}{K. Tretyakov}, \bibinfo{person}{A.
  A.-Cervantes}, \bibinfo{person}{L. G.-Banuelos}, \bibinfo{person}{J. Vilo},
  {and} \bibinfo{person}{M. Dumas}.} \bibinfo{year}{2011}\natexlab{}.
\newblock \showarticletitle{{Fast Fully Dynamic Landmark-based Estimation of
  Shortest Path Distances in Very Large Graphs}}. In \bibinfo{booktitle}{{\em
  CIKM}}.
\newblock


\bibitem[\protect\citeauthoryear{Vaquero, Cuadrado, Logothetis, and
  Martella}{Vaquero et~al\mbox{.}}{2013}]%
        {VCLM13}
\bibfield{author}{\bibinfo{person}{L. Vaquero}, \bibinfo{person}{F. Cuadrado},
  \bibinfo{person}{D. Logothetis}, {and} \bibinfo{person}{C. Martella}.}
  \bibinfo{year}{2013}\natexlab{}.
\newblock \showarticletitle{{Adaptive Partitioning for Large-scale Dynamic
  Graphs}}. In \bibinfo{booktitle}{{\em SOCC}}.
\newblock


\bibitem[\protect\citeauthoryear{Wang, Xiao, Shao, and Wang}{Wang
  et~al\mbox{.}}{2014}]%
        {WXSW14}
\bibfield{author}{\bibinfo{person}{L. Wang}, \bibinfo{person}{Y. Xiao},
  \bibinfo{person}{B. Shao}, {and} \bibinfo{person}{H. Wang}.}
  \bibinfo{year}{2014}\natexlab{}.
\newblock \showarticletitle{{How to Partition a Billion-Node Graph}}. In
  \bibinfo{booktitle}{{\em {ICDE}}}.
\newblock


\bibitem[\protect\citeauthoryear{Wei, Xia, Sha, Xu, He, and Zhou}{Wei
  et~al\mbox{.}}{2013}]%
        {WXSXHZ13}
\bibfield{author}{\bibinfo{person}{J. Wei}, \bibinfo{person}{F. Xia},
  \bibinfo{person}{C. Sha}, \bibinfo{person}{C. Xu}, \bibinfo{person}{X. He},
  {and} \bibinfo{person}{A. Zhou}.} \bibinfo{year}{2013}\natexlab{}.
\newblock \showarticletitle{{Workload-Aware Cache for Social Media Data}}. In
  \bibinfo{booktitle}{{\em APWeb}}.
\newblock


\bibitem[\protect\citeauthoryear{Yang, Yan, Zong, and Khan}{Yang
  et~al\mbox{.}}{2012}]%
        {YYZK12}
\bibfield{author}{\bibinfo{person}{S. Yang}, \bibinfo{person}{X. Yan},
  \bibinfo{person}{B. Zong}, {and} \bibinfo{person}{A. Khan}.}
  \bibinfo{year}{2012}\natexlab{}.
\newblock \showarticletitle{{Towards Effective Partition Management for Large
  Graphs}}. In \bibinfo{booktitle}{{\em SIGMOD}}.
\newblock


\bibitem[\protect\citeauthoryear{Zhao, Sala, Wilson, Zheng, and Zhao}{Zhao
  et~al\mbox{.}}{2010}]%
        {ZSWZZ10}
\bibfield{author}{\bibinfo{person}{X. Zhao}, \bibinfo{person}{A. Sala},
  \bibinfo{person}{C. Wilson}, \bibinfo{person}{H. Zheng}, {and}
  \bibinfo{person}{B.~Y. Zhao}.} \bibinfo{year}{2010}\natexlab{}.
\newblock \showarticletitle{{Orion: Shortest Path Estimation for Large Social
  Graphs}}. In \bibinfo{booktitle}{{\em WOSN}}.
\newblock


\bibitem[\protect\citeauthoryear{Zheng, Labrinidis, and Chrysanthis}{Zheng
  et~al\mbox{.}}{2016}]%
        {ZLC16}
\bibfield{author}{\bibinfo{person}{A. Zheng}, \bibinfo{person}{A. Labrinidis},
  {and} \bibinfo{person}{P.~K. Chrysanthis}.} \bibinfo{year}{2016}\natexlab{}.
\newblock \showarticletitle{{Planar: Parallel Lightweight Architecture-Aware
  Adaptive Graph Re-partitioning}}. In \bibinfo{booktitle}{{\em ICDE}}.
  \bibinfo{pages}{121--132}.
\newblock


\end{thebibliography}
